\documentclass[pdftex,twocolumn,epjc3]{svjour3}
\usepackage{graphics}
\usepackage{amssymb,amsfonts,amsmath} 
\usepackage{graphicx}
\usepackage{cuted}
\usepackage{color}
\usepackage{hyperref}
\usepackage{placeins}
\usepackage{tikz}
\DeclareUnicodeCharacter{0308}{ü}
\usepackage{braket}
\usepackage[
  backend=biber,
  style=phys,
  autocite=inline,
  biblabel=brackets,
  eprint=true,
]{biblatex}
\bibliography{main.bbl}

\tikzset{->-/.style={decoration={
  markings,
  mark=at position .6 with {\arrow{stealth}}},postaction={decorate}}}

\journalname{Eur. Phys. J. B}
\usetikzlibrary{arrows,shapes,positioning}
\usetikzlibrary{decorations.markings}
\usetikzlibrary{decorations.pathmorphing}
\usetikzlibrary{decorations.pathreplacing}
\usetikzlibrary[arrows]
\tikzstyle{arrowstyle}=[scale=1]
\usepackage{tkz-euclide}
\newlength{\overwritelength}
\newlength{\minimumoverwritelength}
\newcommand{\overwrite}[3]{%
  \settowidth{\overwritelength}{$#1$}%
  \ifdim\overwritelength<\minimumoverwritelength%
    \setlength{\overwritelength}{\minimumoverwritelength}\fi%
  \stackrel
    {%
      \begin{minipage}{\overwritelength}%
        \color{#2}\centering\small #3\\%
        \rule{1pt}{9pt}%
      \end{minipage}}
    {\colorbox{#2!30}{\color{black}$\displaystyle#1$}}}

\usetikzlibrary{arrows,shapes,positioning}
\usetikzlibrary{decorations.markings}
\usetikzlibrary{decorations.pathmorphing}
\usetikzlibrary{decorations.pathreplacing}

\usetikzlibrary[arrows]
\usetikzlibrary{fadings,shapes.arrows,shadows}
\tikzstyle{arrowstyle}=[scale=1]
\newcommand{\bvec}[1]{\boldsymbol #1}
\begin{document}
\title{TU$^2$FRG - a scalable approach for truncated unity functional renormalization group in generic fermionic models}
\author{Jonas B.~Hauck\thanksref{addr1}  \and Dante M.~Kennes\thanksref{addr1, addr2}
}                     
\institute{Institute for Theoretical Solid State Physics,
  RWTH Aachen University, 52062 Aachen, Germany and JARA - Fundamentals of Future Information Technology \label{addr1}\and
Max Planck Institute for the Structure and Dynamics of Matter, Center for Free Electron Laser Science, Hamburg, Germany\label{addr2}}
\date{Received: 15.01.2022 / Revised version: date}

\abstractdc{
{Describing the emergence of phases of condensed matter is one of the central challenges in physics. For this purpose many numerical and analytical methods have been developed, each with their own strengths and limitations. The functional renormalization group is one of these methods bridging between efficiency and accuracy. In this paper we derive a new truncated unity (TU) approach unifying real- and momentum space TU, called TU$^2$FRG. This formalism significantly improves the scaling compared to conventional momentum (TU)FRG when applied to large unit-cell models and models where the translational symmetry is broken.}
} 
\maketitle
\section{Introduction}
\begin{figure*}[!htb]
    \centering
    \includegraphics[width = 0.24\linewidth]{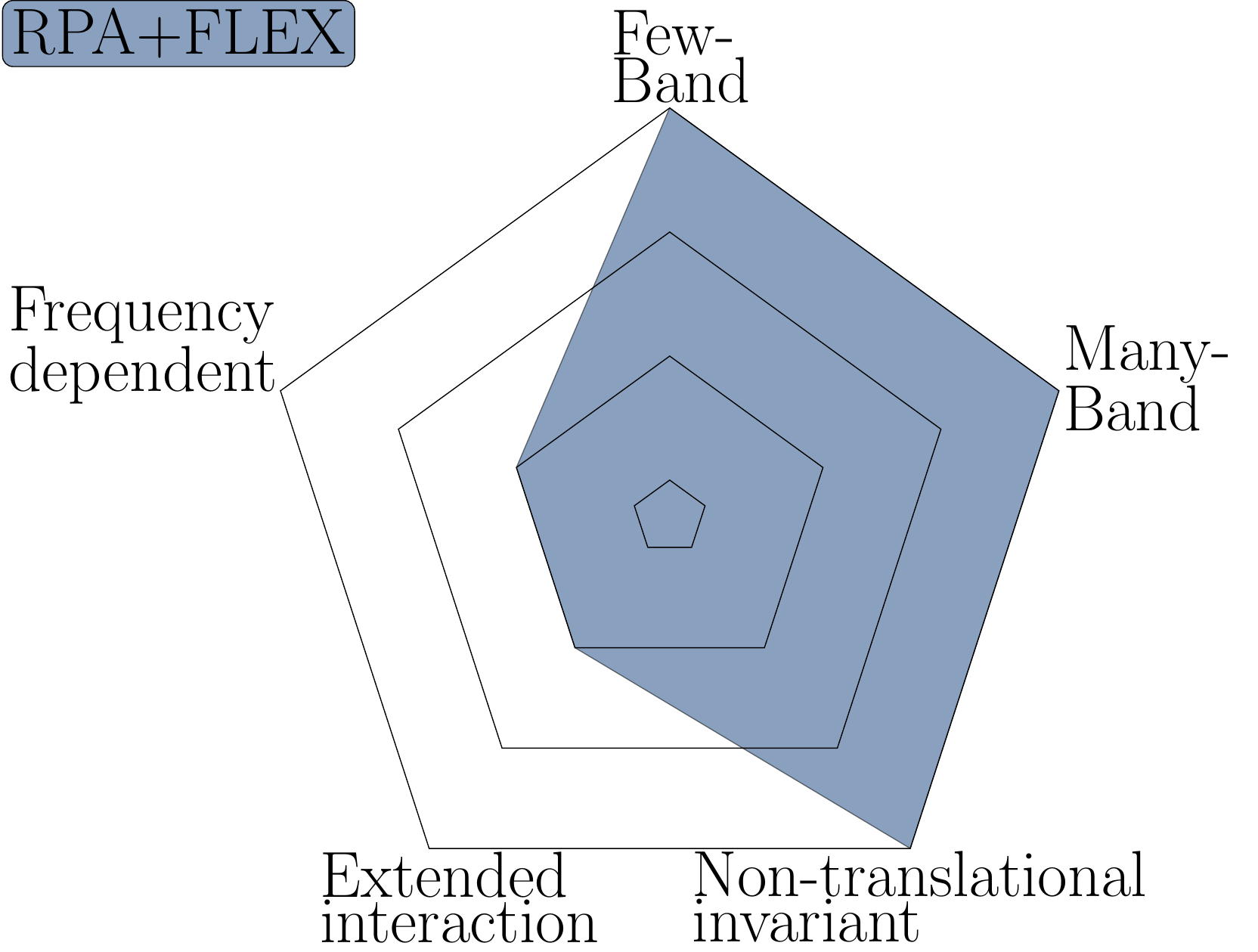}
    \includegraphics[width = 0.24\linewidth]{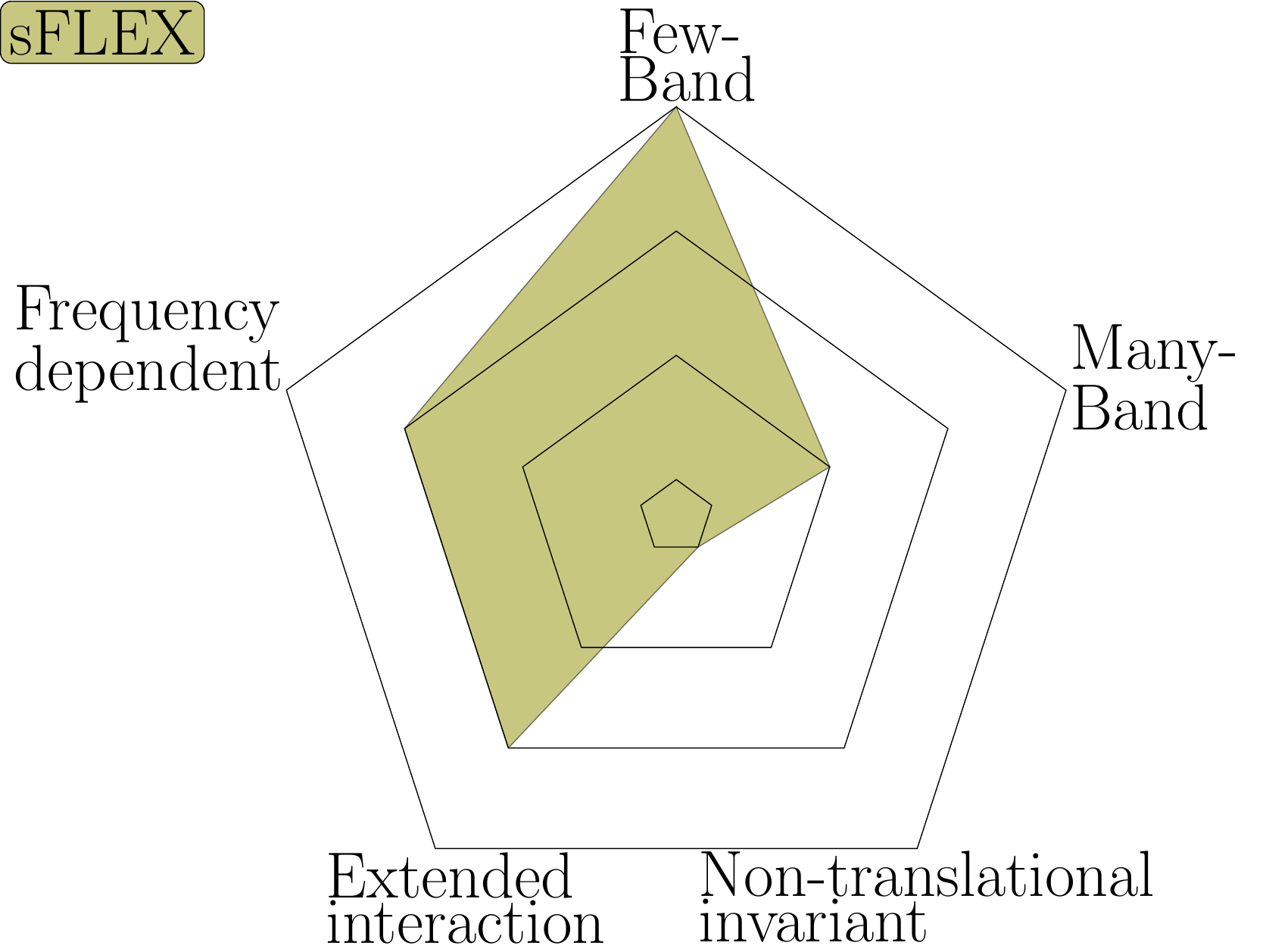}
    \includegraphics[width = 0.24\linewidth]{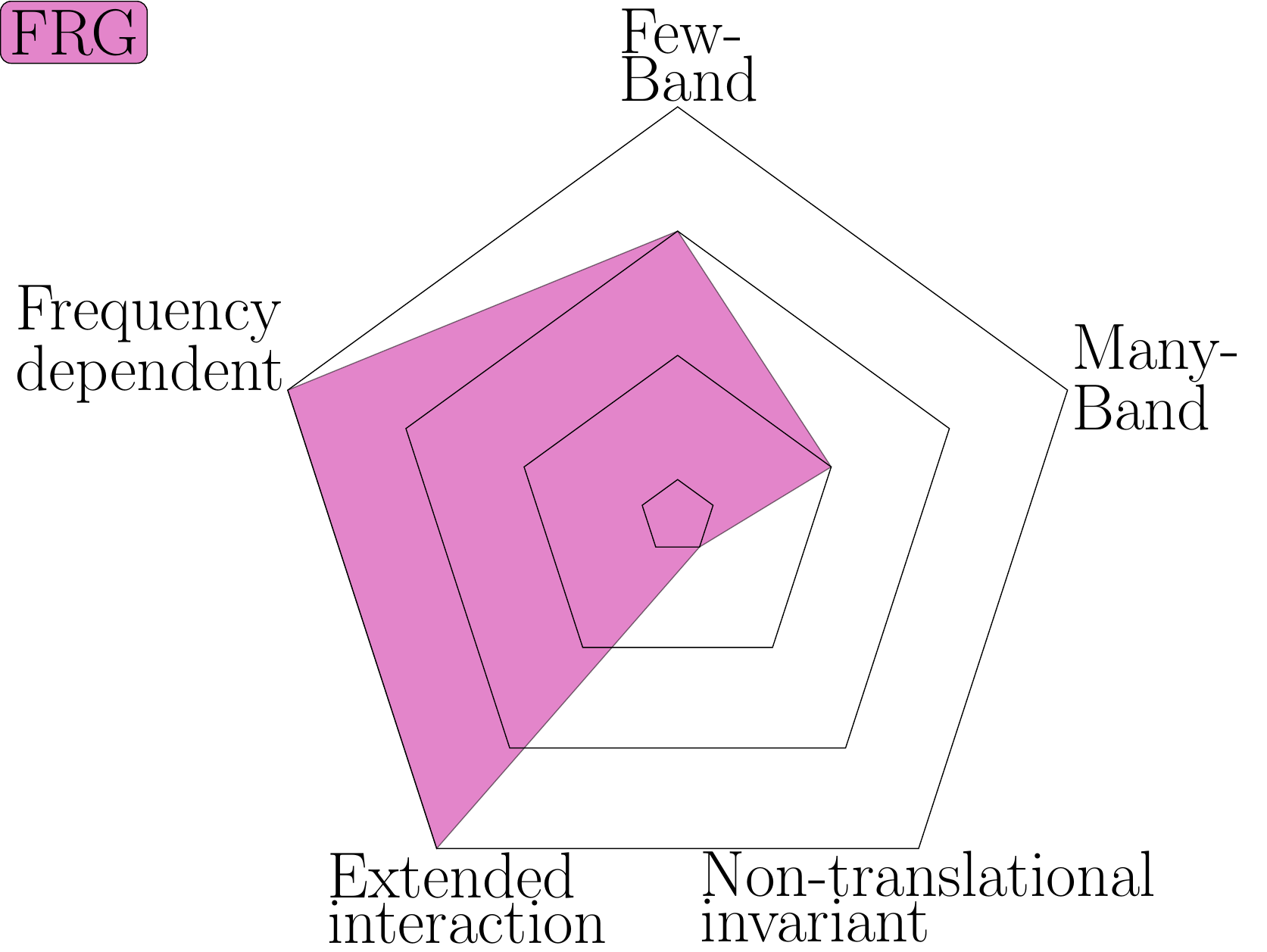}
    \includegraphics[width = 0.24\linewidth]{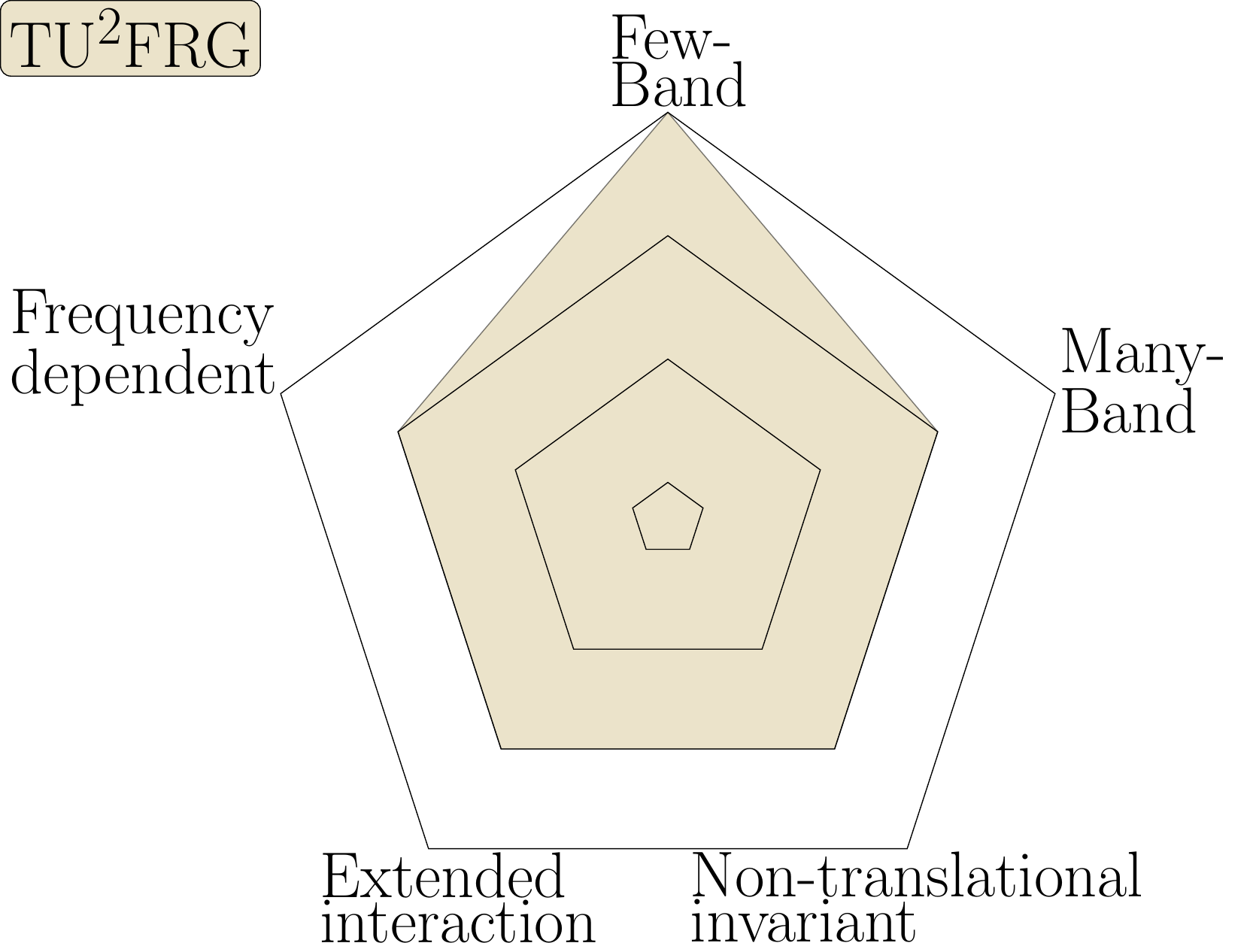}
    \caption{Illustrative comparison of the capabilities of four different methods concerning different use cases. We compare the simple RPA+FLEX without self consistency, the self-consistent FLEX, standard FRG and the here presented TU$^2$FRG. We define few-bands as the range between one and twenty bands and many-band as all systems with more than $20$ bands. { The corner frequency dependence is measured by the number of frequencies included in the method. The capability of each method to process effects of long range interactions is measured in the number of dependencies allowed for the effective interaction generated by the method. For all corners we used the scaling combined with the accuracy in terms of the number of diagrams included by the method to estimate its capability in this direction.}}
    \label{fig:comparision_of}
\end{figure*}

 Predicting the phase diagrams of real materials is one of the central goals and challenges of condensed matter research. For this purpose, many numerical techniques have been developed for many different scenarios~\cite{PhysRev.140.A1133,Schollwoeck,RevModPhys.68.13,PhysRevD.24.2278,Multimessengerpaper}. 
In the weak to intermediate coupling regime, pertubatively motivated approaches have been very successful~\cite{björnson-2021-flex,fischer_spin-fluctuation-induced_2020,Hisashi_Kondo1998,PhysRevB.60.14585,Drchal_2004}: Starting from simple random-phase approximation (RPA)~\cite{PhysRev.112.1900} to the more elaborate self-consistent fluctuation-exchange (FLEX) formalism~\cite{BICKERS1989206}, these methods have been widely used to study magnetism and superconductivity. The biggest drawback of these methods is that they are biased. In particular they do not account for all diagrammatic contributions up to a certain order. This problem can be remedied by adopting the Parquet approximation or using the functional renormalization group (FRG)~\cite{metzner_functional_2012,doi:10.1063/1.1704064}. These methods sum up all diagrammatic contributions up to a certain order, therefore giving a coherent and unbiased picture of the phases of matter. This higher accuracy comes at a higher numerical cost, making calculations of many relevant toy models of real solids hard or even impossible. {Thus, it is critical to} find suitable approximations to reduce the numerical cost of these methods, while maintaining the advantage of unbiased predictions. An illustrative comparison of the three methods discussed above and the method presented in this paper is given in Fig.~\ref{fig:comparision_of}. {There we summarize the capabilities of each of the four methods in five different categories: Few-Band, Many-Band, Non-translational invariant, Extended interactions and frequency dependence. 
Each of the methods has its strengths and weaknesses. For example, plain FRG can be used to consider the full frequency dependence of the effective interaction, but is hardly applicable to non-translationally invariant models without any further approximations. While this can be solved by using the TU$^2$FRG, we lose part of the full frequency dependence to be applicable to a wider set of models as otherwise numerical cost would be too high. RPA is only calculating the spin-fluctuation mediated effective interaction and is thus numerically cheap; but inter-channel feedback is neglected and the  concept of renormalization is not included. sFLEX builds on top of RPA and iterates its self-energy till convergence while including all spin and charge fluctuation diagrams, but this comes at the cost of comparable scaling to FRG for many-bands and non-translational invariant models.}

One approach recently put forward for this purpose is the use of truncated unities, the so called truncated unity FRG (TUFRG)~\cite{lichtenstein_functional_2018} or truncated unity Parquet approach~\cite{eckhardt_truncated_2020}. The main proposition of these approaches is to reduce the numerical effort by reducing the degrees of freedom under consideration by truncating the dependence of weakly varying variables. This approach was already successfully applied in the square lattice Hubbard model~\cite{lichtenstein_high-performance_2017,ehrlich_functional_2020,eckhardt_truncated_2020} and graphene~\cite{de_la_pena_competing_2017,de_la_pena_antiferromagnetism_2017,o_competing_2021,o_effect_2019}.
On the other hand, in the real-space representation the so called extended coupled ladder approximation and analogously the real-space TUFRG  was put forward~\cite{bauer_functional_2014,weidinger_functional_2017,markhof_detecting_2018, hauck_electronic_2021}.
Another big advantage of methods like Parquet or FRG is that in principle one can benchmark the numerical accuracy by taking into account higher order terms. This has been recently implemented in one incarnation in the multiloop-formalism~\cite{kugler-2018-multiloop,tagliavini-2019-convergence,hille-2020-quantitative}, where the authors were able to match results to numerically exact methods in the half-filled Hubbard model up to intermediate interactions strengths.
Thus, there are currently two different directions of research in the methodological development of FRG, the first is the increase of numerical accuracy, the second is the adaption to broader classes of models. In this paper we will present a new approach which falls into the second category. It aims to enable simulations of models with broken translational symmetries, such as quasicrystals~\cite{hauck_electronic_2021}, boundary effects, finite size effects~\cite{hauck_strong_2021} or disorder.
Additionally, we fuse this approach with known momentum-space TUFRG in order to enable studies of models with large unit cells or many orbitals, as in the case of twisted materials or Kagomé metals~\cite{kennes_moire_2021,Ye_2018_kagome_metal}.
For this purpose we derive a full TUFRG-scheme including momentum-, frequency- and real-space unities, in the following called TU$^2$FRG. This can be seen as a key stepping-stone to establish FRG as the default method for calculating electronic instabilities for wide classes of 
electronic models.

\section{Derivation of the flow equations in the unity-space}
We start the derivation from the general flow equations in the second order truncation. As the derivation of these flow equations is well documented~\cite{honerkamp-salmhofer-2001,metzner_functional_2012} we omit it here for the sake of conciseness, and just derive their representation in full unity-space, inserting a unity in all relevant degrees of freedom except spin. The flow equation for the self-energy reads
\begin{align}
\frac{{\rm d}\Sigma(1, 3)}{{\rm d}\Lambda}  =- T\sum S_{2,4}\Gamma(1,2;3,4)e^{i \omega_{2} 0^{+}}, 
\label{eq:selfenergy}
\end{align}
where we defined the single-scale propagator ${S = G\left[\partial_{\Lambda}(G_0)^{-1} \right]G}$ with the full Greens-function defined as {$G^{\Lambda} = \frac{R(\Lambda)}{i\omega - H - R(\Lambda)\Sigma^{\Lambda}} = \left[ (G_0)^{-1} - \Sigma^{\Lambda}\right]^{-1}$}. The flow equation for the effective interaction $\Gamma$ is separable into the three two-particle-irreducible (2-PI) channels according to Eq.~\eqref{eq:2pisep}, each of which is associated with a specific fermionic-bilinear:
 \begin{align}
\Gamma(1,2;3,4) = &U(1,2;3,4) + \Phi^P(1,2;3,4) \nonumber
\\
+&\Phi^D(1,2;3,4)+\Phi^C(1,2;3,4). \label{eq:2pisep}
\end{align}
We have the particle-particle channel ($P$), whose bilinear is of cooper pair type. We have the direct particle hole channel ($D$), which has {a} density-density type bilinear and we have the crossed particle hole channel ($C$) with a spin-spin bilinear. Each of these bilinears can be linked to a different mean-field decoupling, thus, divergences in a channel indicate a phase transition to a certain ordered state~\cite{honerkamp_magnetic_2001}.

This decomposition allows a separation of the flow equations into the three channels~\cite{metzner_functional_2012}:
\begin{subequations} 
\begin{align}
\frac{{\rm d}\Phi^P(1,2;3,4)}{{\rm d}\Lambda} = - \Gamma(1,2;1',2')&\left(G(1';3')S(2';4') \right)\nonumber \\ \cdot\Gamma(3',4';3,4)&, \label{eq::gamma4a} \\
\nonumber \frac{{\rm d}\Phi^C(1,2;3,4)}{{\rm d}\Lambda} = \phantom{-}  \Gamma(1,4';1',4)&L(1',2';3',4')\\
                       \cdot \Gamma(3',2;3, 2')&,  \label{eq::gamma4b}\\
\nonumber \frac{{\rm d}\Phi^D(1,2;3,4)}{{\rm d}\Lambda} = -\Gamma(1,4';3,1')&L(1',2';3',4') \\
                                               \cdot \Gamma(3',2;2', 4)&, 
\label{eq::gamma4c}
\end{align}
\end{subequations}
where in- and out-going indices (of a diagramatic representation of these equations) are separated by a semicolon and we defined ${L(1',2';3',4') = G(1';3')S(4';2') + S(1';3')G(4';2')}$.

In principle, one just needs to integrate Eq.~\eqref{eq:selfenergy} and Eqs.~(\ref{eq::gamma4a},~\ref{eq::gamma4b},~\ref{eq::gamma4c}) numerically, which is a computationally heavy task and is not possible for many models of interest.
In particular, the flow equations computation time scales $\propto \mathcal{O}(N_o^6 N_k^3 N_{\omega}^3)$ with $N_o$ the number of orbitals, sites and spins per unit cell, $N_k$ the number of momentum points and $N_{\omega}$ the number of Matsubara frequencies. In addition, storing the vertex requires $N_o^4 N_k^3 N_{\omega}^3$ elements, such that it becomes apparent that a more efficient representation has to be found for the study multi-orbital/multi-site models.

In the following we will derive the flow equations in the full unity-space. We will argue why this representation is advantageous when compared with brute force solutions, and what its limitations are. To keep our derivation as general as possible, we will keep all dependencies of the vertex; site, momentum, frequency and spin (note that we exclude the case of non-energy-conserving models).
The starting point for our derivation are the Mandelstamm variables, see Eq.~\eqref{Eq::bos_freq}, each of which is associated with one of the channel's momentum transfer. For brevity we condense the momentum and frequency contributions into a four vector denoted as $q$ and collapse the spin and site indices into orbital indices $o$.
\begin{align}
\begin{split}
q_P = k_1 +k_{2} = k_{4} +k_{3}, \\
q_C =  k_{1} - {k}_{4} = {k}_{3}-{k}_{2},\\
q_D = {k}_{1} - {k}_{3} = {k}_{4} -{k}_{2}.
\end{split}
\label{Eq::bos_freq}
\end{align}
These Mandelstamm variables can now be used to rewrite the flow equations in  compact fashion. For brevity we introduce an abbreviation for sets of increasing indices $i_1,i_2,i_3,i_4$ as $i_1.._4$. Additionally we use an altered Einstein sum convention, where we sum or integrate out each doubly occurring index. {It has to be kept in mind that the momentum summations stem from a Fourier transformation, therefore they always include a normalization factor, which too gets suppressed for brevity.}
Thereby we arrive at the following set of flow equations.
\begin{subequations} 
\begin{align}
 &\frac{{\rm d}\Sigma_{o_1, o_2}( {k})}{- T{\rm d}\Lambda}  =S_{o_1',o_2'}({k'})
 \Gamma_{o_1,o_1',o_2,o_2'}( {k},{k'},{k}) e^{i \omega' 0^{+}}, 
 \\
\nonumber &\frac{{\rm d}\Phi^P_{o_1.._4}({k}_1.._3)}{{\rm d}\Lambda} = -\frac{1}{2} \Gamma_{o_1, o_2;o_1', o_2'}({k}_1,{k}_2;{q})
\\\nonumber & \qquad\qquad \qquad \qquad [G_{o_1';o_3'}({q})S_{o_2';o_4'}({k}_1+{k}_2-{q})
\\\nonumber & \qquad\qquad \qquad \qquad+ S_{o_1';o_3'}({q})G_{o_2';o_4'}({k}_1+{k}_2-{q})] 
\\  & \qquad\qquad \qquad \qquad\Gamma_{o_3', o_4';o_3, o_4}({q},{k}_1+{k}_2-{q};{k}_3)
, \\
\nonumber &\frac{{\rm d}\Phi^C_{o_1.._4}({k}_1.._3)}{{\rm d}\Lambda} =  \phantom{-}\Gamma_{o_1,o_4',o_1',o_4}({k}_1,{k}_2-{k}_3+{q};{q})
\\\nonumber & \qquad\qquad \qquad \qquad [G_{o_1';o_3'}({q})S_{o_4';o_2'}({k}_2-{k}_3+{q}) \\\nonumber & \qquad \qquad \qquad \qquad+ S_{o_1';o_3'}({q})G_{o_4';o_2'}({k}_2-{k}_3+{q}) ]
\\ & \qquad\qquad \qquad \qquad \Gamma_{o_3',o_2;o_3,o_2'}({q},{k}_2;{q})
,\\
\nonumber &\frac{{\rm d}\Phi^D_{o_1.._4}({k}_1.._3)}{{\rm d}\Lambda} =  -\Gamma_{o_1, o_4';o_3, o_1'}({k}_1,{q}-{k}_1+{k}_3 ;{k}_3) 
\\\nonumber & \qquad\qquad \qquad \qquad[G_{o_1';o_3'}({k}_3-{k}_1+{q})S_{o_4';o_2'}({q})
\\ \nonumber & \qquad\qquad \qquad \qquad + S_{o_1';o_3'}({k}_3-{k}_1+{q})G_{o_4';o_2'}({q})]
\\ & \qquad\qquad \qquad \qquad\Gamma_{o_3', o_2;o_2', o_4}({q},{k}_2;{q}-{k}_1+{k}_3).
\end{align}
\end{subequations}
These equations are the starting point for all subsequent derivations. So far, spin and orbitals or sites are treated equally, this means that $o_i = (\tilde{o}_i,s_i)$ is a multi-index consisting of both. {For the following derivation we will denote spin and orbitals separately as spins and orbitals need to be treated differently and redefine $\tilde{o} \equiv o$ as the site and orbital index. }

The motivation to transform into the unity-space follows from the behavior of RPA-like resummations of each channel~\cite{lichtenstein_high-performance_2017}; In the flow equations, the lowest order of diagrams we do not account for is $U^3$, which fixes our truncation order in the interaction. Usually the initial interaction has a finite range, or drops off as a function of the distance. Therefore, for the sake of the argument we now assume an interaction between nearest neighbors, with a density-density bilinear, thus $U_{o_1,o_2,o_3,o_4}^{s_1,s_2,s_3,s_4}(\omega_1,\omega_2;\omega_3) = U_0 \delta_{o_1,o_3}^{s_1,s_3}\delta_{o_2,o_4}^{s_2,s_4} \delta_{<o_1,o_2>}$, where $\delta_{<o_1,o_2>}$ is only one if the two indices belong to neighbouring sites. If we now insert this interaction into the $P$-channel flow equation we obtain
\begin{align}
    \nonumber \frac{{\rm d}\Phi_{o_1.._4}^{P|s_1.._4}(\omega_1,\omega_2;\omega_P)}{{\rm d}\Lambda} =& \sum_{\omega} U_0^2 \delta_{<o_1,o_2>} \delta_{<o_3,o_4>}  \\ &\nonumber[G_{o_1;o_3}^{s_1;s_3}(\omega)S_{o_2;o_4}^{s_2;s_4}(\omega_P-\omega) \\ &+S_{o_1;o_3}^{s_1;s_3}(\omega)G_{o_2;o_4}^{s_2;s_4}(\omega_P-\omega)].
\end{align}
We observe that in the frequency space we do not generate terms depending on $\omega_1$ and $\omega_2$, whereas in real-space, we will keep on generating terms $\propto \delta_{<o_1,o_2>} \delta_{<o_3,o_4>}$ if we insert $\Phi^P$ into the right hand side again. Thus, on {a} single channel level, we cover all dependencies by only allowing for very specific index combinations, and neglecting all others as they will be always zero. The argument carries over to  the other two channels. If we now include the feedback in between the channels by reconstructing $\Gamma$ we will again generate additional dependencies. These will have a hierarchy in the distance, where larger distances correspond to higher order interaction terms increasing with the distance from the initial index combinations. As a consequence, we can identify leading and subleading dependencies of each channel, which can be exploited to arrive at an efficient description of the problem.

For this purpose we define a set of functions $g_{b_y}(o_x,{k})$ which give for each site $x$ the momentum and frequency dependent connections to site $y$. These functions are required to form an orthonormal basis on the space of the momentum, site and frequency, thus fulfilling 
\begin{align}
&\sum_{b_1} g_{b_1}(o_2, k)g_{b_1}^*(o_3, k') =  \delta_{ k, k'}\delta_{o_2,o_3},\\
&\sum_{o, k} g_{b_1}(o, k)g_{b_2}^*(o, k) = \delta_{b_1,  b_2}.
\end{align} 
The choice of the basis set is not unique and we will discuss two possibilities later. With these unities we can define projections onto the leading dependencies of each channel as
\begin{align}
\hat{P}[\Gamma]&_{o_1,o_3}^{b_{1},b_3}( q_P)_{s_1;s_3}^{s_2;s_4} = \int d k_1\, d {k}_3\, \sum_{o_2,o_4} g_{b_1}(o_2, k_1) \nonumber 
\\ &\quad \cdot g^*_{b_3}(o_4, k_3)\Gamma_{o_1.._4}( k_1, q_P- k_1; k_3)_{s_1.._4}, \\
\hat{C}[\Gamma]&_{o_1,o_3}^{b_{1},b_3}( q_C)_{s_1;s_3}^{s_4;s_2} = \int d k_1\, d {k}_3\, \sum_{o_2,o_4} g_{b_1}(o_4,  k_1)\nonumber 
\\ &\quad \cdot g^*_{b_3}(o_2, k_3)\Gamma_{o_1.._4}( k_1, k_1- q_C; k_3)_{s_1.._4}, \\
\hat{D}[\Gamma]&_{o_1,o_4}^{b_{1},b_4}( q_D)_{s_1;s_4}^{s_3;s_2} = \int d k_1\,d {k}_4\, \sum_{o_2,o_3} g_{b_1}(o_3, k_1)\nonumber 
\\ &\quad \cdot g^*_{b_4}(o_2, k_4) \Gamma_{o_1.._4}( k_1, k_4- q_D; k_1- q_D)_{s_1.._4}.
\end{align}
Note the spin reordering in each of the projections, which is performed to enable a reformulation of the flow equations in terms of batched matrix multiplications. These projections are designed such that we fully keep all ladder-like contributions of each channel. Depending on how we truncate the basis we take the feedback in between the channels into account only approximately.
The inverse projections follow from the completeness relation as 

\begin{align}
\hat{P}^{-1}[\hat{P}[&\Gamma]]_{o_1.._4}( k_1.._3)_{s_1.._4} =  \sum_{b_1,b_3} g^*_{b_1}(o_2, k_1)g_{b_3}(o_4, k_3)\nonumber 
\\ &\cdot \hat{P}[\Gamma]_{o_1,o_3}^{b_1,b_3}( q_P)_{s_1;s_3}^{s_2;s_4} , \\
\hat{C}^{-1}[\hat{C}[&\Gamma]]_{o_1.._4}( k_1.._3)_{s_1.._4} = \sum_{b_1,b_3}  g^*_{b_1}(o_4, k_1)g_{b_3}(o_2, k_3)\nonumber 
\\ &\cdot \hat{C}[\Gamma]_{o_1,o_3}^{b_1,b_3}( q_C)_{s_1;s_3}^{s_4;s_2}, \\
\hat{D}^{-1}[\hat{D}[&\Gamma]]_{o_1.._4}( k_1.._3)_{s_1.._4} = \sum_{b_1,b_4} g^*_{b_1}(o_3, k_1)\nonumber 
\\ &\cdot g_{b_4}(o_2,  k_1+ k_2- k_3) \hat{D}[\Gamma]_{o_1,o_4}^{b_1,b_4}( q_D)_{s_1;s_4}^{s_3;s_2} .
\end{align}
{To benefit from the interaction hierarchy in the basis function dependence, we truncate the unity to small number of basis functions. Thereby, the projection procedure becomes approximate and the full effective interactions cannot be recovered exactly anymore. While the error between the truncated and the full unity is uniform in the number of basis functions included, the error in the projected channels is non-uniform, see discussion above. This allows for a faithful representation of the effective interaction with only a few basis functions.}
For better readability we explicitly expand the summations when we insert a new unity. We begin with the flow equations for $\hat{P}[\Phi^P] \equiv P$, starting from Eq.~\eqref{eq::gamma4a} in Eq.~\eqref{eq:proj_P_flow_b}.

\begin{figure*}[t!]
\begin{align}
\frac{dP_{o_1,o_3}^{b_{1},b_3}( q_P)_{s_1;s_3}^{s_2;s_4}}{{\rm d}\Lambda} &=  - g_{b_1}(o_2, k_1)g^*_{b_3}(o_4, k_3)\frac{1}{2} \Gamma_{o_1, o_2;o_1', o_2'}^{s_1,s_2;s_1',s_2'}( k_1, q_P- k_1; p)\nonumber
\\ 
&\qquad\quad  \label{eq:proj_P_flow_b} \left(G_{o_1';o_3'}^{s_1',s_3'}({ p})S_{o_2';o_4'}^{s_2',s_4'}( q_P-{ p}) + G\leftrightarrow S) \right) \Gamma_{o_3', o_4';o_3, o_4}^{s_3',s_4';s_3,s_4}({ p}, q_P-{ p}; k_3)
\\
&\quad=  \nonumber - g_{b_1}(o_2, k_1)g^*_{b_3}(o_4, k_3)\frac{1}{2} \Gamma_{o_1, o_2;o_1', o_2'}^{s_1,s_2;s_1',s_2'}( k_1, q_P- k_1; p)
\\
&\qquad\quad \nonumber  \int d p_1\,\sum_{n_2} \int d  p_2 \,\sum_{n_4} \delta( p- p_1)\delta_{o_2',n_2}\left(G_{o_1';o_3'}^{s_1',s_3'}({ p_1})S_{n_2;n_4}^{s_2',s_4'}( q_P-{ p_1}) + G\leftrightarrow S) \right) 
\\
&\qquad\quad   \delta_{o_4',n_4} \delta( p_2- p_1)\Gamma_{o_3', o_4';o_3, o_4}^{s_3',s_4';s_3,s_4}({ p_2}, q_P-{ p_2}; k_3)
\\
&\quad= \nonumber  -g_{b_1}(o_2, k_1)\frac{1}{2} \Gamma_{o_1, o_2;o_1', o_2'}^{s_1,s_2;s_1',s_2'}( k_1, q_P- k_1; p) g^*_{b_1'}(o_2', p)
\\ 
& \qquad\quad \nonumber  g_{b_1'}(n_2, p_1)g^*_{b_3'}(n_4, p_1)\left(G_{o_1';o_3'}^{s_1',s_3'}({ p_1})S_{n_2;n_4}^{s_2',s_4'}( q_P-{ p_1}) + G\leftrightarrow S) \right) 
\\ 
& \qquad\quad   g_{b_3'}(o_4, p_2) \Gamma_{o_3', o_4';o_3, o_4}^{s_3',s_4';s_3,s_4}({ p_2}, q_P-{ p_2}; k_3)g^*_{b_3}(o_4, k_3) 
\\
&\quad= -\frac{1}{2} \hat{P}[\Gamma]_{o_1,o_1'}^{b_{1},b_{1}'}( q_P)_{s_1;s_1'}^{s_2;s_2'} \, L^{pp;b_{1}',b_{3}'}_{o_1',o_3'}( q_P)_{s_1';s_3'}^{s_2';s_4'}\, \hat{P}[\Gamma]_{o_3',o_3}^{b_{3}',b_{3}}( q_P)_{s_3';s_3}^{s_4';s_4}.
\label{eq:proj_P_flow}
\end{align}
\end{figure*}
Here, we defined the particle-particle loop derivative and analogously the particle-hole loop derivative as
\begin{align}
L^{pp;b_1,b_3}_{o_1,o_3}&( q_P)_{s_1,s_3}^{s_2,s_4} = \int d p_1 \, g_{b_1}(n_2, p_1)g^*_{b_3}(n_4, p_1) \nonumber
\\ &\cdot\left(G^{s_1,s_3}_{o_1;o_3}(p_1)S^{s_2,s_4}_{n_2;n_4}( q_P- p_1)+ G\leftrightarrow S) \right), \label{eq:ppl} \\
L^{ph;b_{1},b_{3}}_{o_1,o_3}&( q_X)_{s_1;s_4}^{s_3;s_2} = \int d p_1 \, g_{b_1}(n_4, p_1)g^*_{b_3}(n_2, p_1) \nonumber
\\ &\cdot\left(G^{s_1,s_3}_{o_1;o_3}({ p_1})S^{s_4,s_2}_{n_4;n_2}({ p_1}- q_X)+ G\leftrightarrow S) \right)\label{eq:phl}, 
\end{align}
where $X\in \{C,D\}$. The derivations for the flow equations for $\hat{C}[\Phi^C]$ and $\hat{D}[\Phi^D]$ follow analogously (as can be seen in App.~\ref{App::floweq}) and result in 

\begin{align}
\frac{{\rm d}\hat{C}[\Phi^C]_{o_1,o_3}^{b_1,b_3}( q_C)_{s_1;s_3}^{s_4;s_2}}{{\rm d}\Lambda} =\phantom{-}&\hat{C}[\Gamma]_{o_1,o_1'}^{b_{1},b_{1}'}( q_C)_{s_1;s_1'}^{s_4;s_4'}\nonumber \\  \quad \cdot  L^{ph;b_{1}',b_{3}'}_{o_1',o_3'}( q_C)_{s_1';s_3'}^{s_4';s_2'}  \, &\hat{C}[\Gamma]_{o_3',o_3}^{b_{3}',b_{3}}( q_C)_{s_3';s_3}^{s_2';s_2},
\label{eq:proj_C_flow}
\\
\frac{{\rm d}\hat{D}[\Phi^D]_{o_1,o_4}^{b_1,b_4}( q_D)_{s_1;s_4}^{s_3;s_2}}{{\rm d}\Lambda} = \nonumber -&\hat{D}[\Gamma]_{o_1,o_1'}^{b_1,b_1'}(q_D)_{s_1;s_1'}^{s_3;s_4'}\\\cdot L^{ph;b_{1}',b_{4}'}_{o_1',o_4'}( q_D)_{s_1';s_3'}^{s_4';s_2'}\, &\hat{D}[\Gamma]_{o_4',o_4}^{b_{4}',b_{4}}( q_D)_{s_3';s_4}^{s_2';s_2}.
\label{eq:proj_D_flow}
\end{align}

So far, this is a mere reformulation, but we already note that the flow equations are transformed into matrix products, and that the summation over momenta is now only required inside the loop derivative calculations.

As it is impossible for many models to store the full reconstructed vertex we instead store the three projected channels $\hat{P}[\Phi^P]\equiv P$, $\hat{C}[\Phi^C]\equiv C$ and $\hat{D}[\Phi^D]\equiv D$. This reduces the memory requirement from {$N_o^4 N_k^3 N_{\omega}^3 N_{s}^4$ to $N_o^2 N_k N_{\omega} N_b^2 N_{s}^4$ (with $N_s$ the number of spins and $N_o$ th number of sites and orbitals in the unit cell)}. Depending on how many basis functions are required to reach convergence of the calculations, this can be a drastic reduction. For the derivation of the unity-space self-energy we recall that in principle we can approximately restore the full vertex as
\begin{align}
\Gamma_{o_.._4}^{s_.._4}(k_1.._3) =\phantom{+} \nonumber \hat{P}^{-1}[P_{o_1,o_2}^{b_1,b_2}(&q_P)_{s_1;s_3}^{s_2;s_4}] \\+ \hat{C}^{-1}[C_{o_1,o_2}^{b_1,b_2}(q_C)_{s_1;s_3}^{s_4;s_2}] +& \hat{D}^{-1}[D_{o_1,o_2}^{b_1,b_2}(q_D)_{s_1;s_4}^{s_3;s_2}]. \label{eq:reconst}
\end{align} 
Inserting this into the self-energy flow equation results in
\begin{align}
\frac{{\rm d}\Sigma^{s_1;s_3}_{o_1;o_3}( k_1)}{-T\cdot d\Lambda} &=S^{s_2;s_4}_{o_2;o_4}( k_2)\nonumber
\\\cdot {\Big[}\hat{P}^{-1}[P]& + \hat{C}^{-1}[C] + \hat{D}^{-1}[D]{\Big]}^{s_1.._4}_{o_1.._4}( k_1, k_2;  k_1)
\\
=\nonumber S^{s_2;s_4}_{o_2;o_4}( &k_2)\Big[
\\g^*_{b_1}(&o_2, k_1)g_{b_3}(o_4, k_1)\hat{P}[\Phi^P]_{o_1,o_3}^{b_{1},b_3}( k_1+ k_2)_{s_1;s_3}^{s_2;s_4}\nonumber \\
\nonumber \quad+ g^*_{b_1}(&o_4, {k}_1)g_{b_3}(o_2, {k}_1)\hat{C}[\Phi^C]_{o_1,o_3}^{b_{1},b_3}( k_1- k_2)_{s_1;s_3}^{s_4;s_2} \\
\quad+ g^*_{b_1}(&o_3, {k}_1)g_{b_4}(o_2, k_2)\hat{D}[\Phi^D]_{o_1,o_4}^{b_{1},b_4}(0)_{s_1;s_4}^{s_3;s_2}\Big].
\label{eq:self_flow}
\end{align}

The FRG in unity-space has therefore reduced to Eqs.~(\ref{eq:proj_P_flow},~\ref{eq:proj_C_flow},~\ref{eq:proj_D_flow},~\ref{eq:self_flow}) which need to be integrated. So far we did not discuss the specific form of the unity, nor specify how we implement these equations.

\section{Specification of the unity}
\label{sec::spec_un}
\begin{figure}[!htb]
    \centering
    \includegraphics[width = 0.9\linewidth]{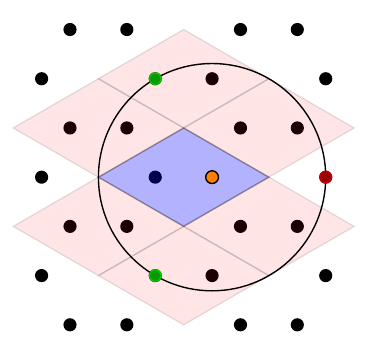}
    \caption{Visualisation of the first form factor shell in a honeycomb lattice (graphene), the reference unit cell is marked in light blue, the six first form factor shells are marked in pink. We observe, that if we include all bonds within the unit cell and the first form factor shell, we still do not include all connections within a distance of $\frac{2}{\sqrt{3}}$, but instead we lack the connection to the site marked in red.}
    \label{fig:bond_issue}
\end{figure}

In this section we will discuss a suitable choice of momentum- and real-space unities. For the frequency unity we will  stick to an on-site form factor expansion in the following, corresponding to a single-frequency-per-channel approximation~\cite{reckling_approximating_2018,weidinger_functional_2017,markhof_detecting_2018}. For a more in depth discussion of the frequency dependence we refer the reader to~{\cite{2012,2017,wentzell-2020-frequency,yirga_frequency_2021}.} To differentiate between the full basis functions we discussed before and the momentum- and real-space basis, we refer to the latter  as form-factor-bonds.
The intuitive way to implement the unity in momentum- and real-space is to utilize a two step procedure: We formally write
\begin{equation}
    g_{b_1}(o_2,\bvec k_1) = \tilde{g}_{\tilde{b}_1}(o_2)\cdot f_m(\bvec k_1),
\end{equation}
which amounts to using completely separate sets of bonds and form-factors. This is easy to implement as it simply adds another projection into existing momentum-space TUFRG codes. Additionally, the projections can be pulled apart making numerical implementations faster and easier to handle.
However, this simple approach comes with a big drawback: For models with multiple sites per unit cell, graphene to name just one example, this approach always breaks the rotational symmetry as soon as we introduce a cutoff distance and neglect corresponding form-factor-bonds, as visualized in Fig.~\ref{fig:bond_issue}.
The problem arises, as the definition of a unit cell introduces a preferred direction for each site, and momentum form-factors are generated in shells around the origin. Thus, the real space distance between sites is not covered correctly in the form-factor shells. Via this inconsistency we take some length-scales, and thereby interaction orders, only partially into account destroying the unbiased nature of the method.
Even though this is not a problem if we converge in form-factor-bonds, it is still a bias and could possibly lead to unexpected behavior and renders observations of nematic phases doubtful.

Luckily, these issues can be resolved. For this, we start with a full real-space description of the lattice, where we set up bonds on the full lattice as real-space Kronecker deltas, as defined in Eq.~\eqref{eq:full_bond}.
\begin{equation}
g_{\tilde{b}_{i}}(\tilde{o}_{j}) = \delta_{\tilde{o}_{i} + \tilde{b}_{i},  \tilde{o}_{j}}  = \delta_{\tilde{\bvec r}_i+\tilde{\bvec b}_i,\tilde{\bvec r}_j}.
\label{eq:full_bond}
\end{equation}
Here $\tilde r$ as opposed to $r$ indicates that the objects are full lattice vectors. Therefore, the bond $\tilde{b}_{i}$ connects the site with index $\tilde{o}_{i}$ to another site, such that only if $\tilde{o}_{j}$ is equivalent to this site, the Kronecker delta is non-zero.

This basis can now be truncated according to the length of the corresponding bonds ensuring the conservation of rotational and inversion symmetries. To return to a mixed space form-factor-bond basis, we need to Fourier transform these bonds according to:
\begin{equation}
g_{b_i}(o_j,\bvec k,\bvec k') = \sum_{\bvec R_{i},\bvec R_{j}} e^{-i\bvec k\bvec R_{i}}e^{-i\bvec k'\bvec R_{j}}  \delta_{\tilde{\bvec r}_i+\tilde{\bvec b}_i,\tilde{\bvec r}_j},
\end{equation}
where we defined $\tilde{\bvec r}_i = \bvec R_{i}+ \bvec r_i$, with $\bvec R_i$ the existing lattice vectors and $\bvec r_i$ the vectors within the unit cell. We also introduce $\tilde{\bvec b_i} = \bvec B_{i}+\bvec b_i$ for the bonds. Again $\bvec b_i$ is equivalent to the connection between site $i$ and the image of another site $j$ within the same unit cell and $\bvec B_{i}$ gives the connection between the unit cells. It is important to notice, that these two length scales are fully separate, which means that two vectors from the two different sets, within the unit cell and beyond the unit cell, can never add to zero.

From the translational invariance it follows that the sum of two lattice vectors must be a lattice vector, i.e.~$\bvec R_j = \bvec R_l+\bvec R_i$ and using this we can write
\begin{align}
g_{b_i}(o_j,&\bvec k,\bvec k') = \sum_{\bvec R_i,\bvec R_j} e^{-i\bvec k\bvec R_{i}}e^{-i\bvec k'\bvec R_{j}}  \delta_{\bvec r_i+\tilde{\bvec b}_i+\bvec R_i,\bvec r_j+\bvec R_j}
\\
&= \sum_{\bvec R_i,\bvec R_l} e^{-i(\bvec k+\bvec k')\bvec R_{i}}e^{-i\bvec k\bvec R_{l}}  \delta_{\bvec r_i+\tilde{\bvec b}_i, \bvec r_j+\bvec R_{l}}
\\
&= \delta_{\bvec k,-\bvec k'} \sum_{\bvec R_{l}} e^{-i\bvec k\bvec R_{l}}  \delta_{\bvec r_i+\bvec b_i+\bvec B_i,\bvec r_j+\bvec R_{l}}.
\end{align}
Within the Kronecker delta we have two different and separate length scales, one within the unit cell, and one beyond. {Due to the fully separate nature of the two length scales, the Kronecker delta can only be non-zero if both $\delta_{\bvec r_i+\bvec b_i,\bvec r_j}$ and $\delta_{\bvec R_{l},\bvec B_i}$ hold, which allows us to split the Kronecker delta into two parts}
\begin{align}
g_{b_i}(o_j,\bvec k) &=  \sum_{\bvec R_{l}} e^{-i\bvec k\bvec R_{l}}  \delta_{\bvec r_i+\bvec b_i,\bvec r_j}\delta_{\bvec R_{l},\bvec B_i}
\\
&= e^{-i\bvec k\bvec B_i}  \delta_{\bvec r_i+\bvec b_i,\bvec r_j}.
\label{}
\end{align}
As one would naively expect  we obtain as form-factor-bonds the plain wave form factor multiplied by a suitable real-space Kronecker-delta, which is also obtained when applying two separate unities for real- and momentum-space after applying the so called filtering process of Ref.~\cite{o_competing_2021}, in which the rotational symmetry gets explicitly enforced.

{For each model, it has to be ensured that the results are converged in form-factor-bonds. These runs are costly in computation time and complex as the convergence is different at each point in parameter space. Thus it is important to understand the physical implications of a certain bond cutoff. For this purpose the pure real space representation of the form-factor-bonds, see Eq.~\eqref{eq:full_bond}, is most suitable. We observe, that removing certain bond vectors $\tilde{\bvec b}_i$ amounts to neglecting certain orbital combinations in the inter-channel projections. Their importance can be deduced by inspecting single-channel flows, which amount to RPA-like resummations.}

The TU$^2$FRG code consists of three numerically challenging steps; the calculation of the loop derivatives (Eq.~\eqref{eq:ppl} and Eq.~\eqref{eq:phl}), the inter-channel projections (Eq.~\eqref{eq:reconst}) and the flow equations~(Eq.~(\ref{eq:proj_P_flow},~\ref{eq:proj_C_flow},~\ref{eq:proj_D_flow},~\ref{eq:self_flow})). We will now discuss each implementation in detail. The flow equations in the representation presented here already have the form of batched matrix products, which are implemented in BLAS-libraries and are numerically very efficient. Thus we only discuss the projections and the loop derivative implementation.

\section{Implementation of the flow equations}
\subsection{Loop derivatives}
The loop derivatives with the above defined form-factor-bonds inserted reduce to the following expressions
\begin{align}
L^{pp;b_1,b_3}_{o_1,o_3}(q_P)_{s_1;s_3}^{s_2;s_4} &= \int dp_1 \, e^{-i\bvec p_1 \bvec B_1} e^{i\bvec p_1 \bvec B_3} \\ 
\cdot \Big(G_{o_1;o_3}^{s_1,s_3}&(p_1)S_{o_1+b_1;o_3+b_3}^{s_2,s_4}( q_P- p_1) + G\leftrightarrow S \Big), \nonumber
\\
L^{ph;b_1,b_3}_{o_1,o_3}(q_X)^{s_4;s_2}_{s_1;s_3} &= \int dp_1 \, e^{-i\bvec p_1 \bvec B_1} e^{i\bvec p_1 \bvec B_3} \\\cdot 
\Big(G_{o_1;o_3}^{s_1,s_3}&(p_1)S_{o_1+b_1;o_3+b_3}^{s_4,s_2}(p_1-q_X) + \nonumber G\leftrightarrow S \Big). 
\end{align}
As the integral over $\bvec p_1$ is decoupled from the flow equation, we can refine the momentum integration of the loop~\cite{lichtenstein_high-performance_2017,honerkamp_breakdown_2001}.
This can either be done by choosing a separate grid for the integration or by implementing an adaptive integration scheme.
There are multiple options to implement this refinement; The simplest option is to use a finer mesh for the $p_1$ integration. This has the advantage of conserving all symmetries but is numerically more demanding then the second strategy.
Alternatively we add an additional sum around each k-point which sums up a mini Wigner-Seitz cell around this momentum point for each single-scale propagator-propagator product, {analogously to the refinment used in $N$-patch FRG schemes~\cite{honerkamp_breakdown_2001}.}

So far we did not specify the cutoff-function we chose for the calculations. Here, many different cutoffs are possible each with specific advantages and disadvantages.
The $\Omega$-cutoff~\cite{husemann_efficient_2009} for example is a smooth cutoff which simplifies the integration of the flow equation in the case of self-energy feedback.
The temperature cutoff~\cite{honerkamp-salmhofer-2001} allows for a physical interpretation of the critical scale as a critical temperature and the interaction cutoff~\cite{honerkamp_interaction_2004} allows for scanning for the critical interaction strength in a single FRG run.
Each of these cutoffs comes with a more or less severe drawback, for example the interaction cutoff does not regularize infrared divergences.
The biggest complications arise if one is interested in large unit cell models, as for example twisted materials~\cite{kennes_moire_2021, klebl_functional_2020,klebl_inherited_2019}.
Here the analytic Matsubara summation scales $\propto N_o^4$, making it the numerical most costly part of the whole calculation by a factor of $N_o$.
On the other hand, the numerical Matsubara summation requires summing up many frequencies for convergence. If the calculation of the Greens-function is non-negligible, this step can also become the bottleneck of the calculation as for a reasonable resolution we need many frequencies, for which in each step the Green-functions have to be recalculated.
A solution to these numerical issues is the sharp cutoff $R(\Lambda)= \theta(|\omega|-\Lambda)$, which reduces the number of Greens-functions required to be calculated in each step to $2N_{\omega}-1$, but introduces discontinuities of the loop derivatives on the frequency axis.
Therefore, the integration of the flow equations has to be performed more carefully, i.e.~the integrator must never integrate over one such discontinuity. {The integration procedure has to be adopted such that we end with an integration step right before the discontinuity and the next step is then performed starting from a point right behind it. Thereby we minimize the numerical error introduced by the discontinuity.}

\subsection{Projection}
\label{ssec:proj}
As we truncate the unity we are unable to exactly recover the three channels $\Phi$ or the effective vertex $\Gamma$ exactly. Additionally due to memory constraints it is often impossible to even store the full vertex.
Therefore, we need to derive efficient formulas for these inter-channel projections. The naive form of these projections, see Eq.~\eqref{eq:simple_proj} requires three Brillouin-zone integrations in addition to four form-factor-bond  sums, making it numerically demanding.
We will instead derive a form which has superior scaling and requires as few operations on the projected vertex channels as possible, as these are the largest objects in our calculation. Note that in pure momentum space, such derivations have been performed similarly~\cite{lichtenstein_high-performance_2017,de_la_pena_antiferromagnetism_2017} and are dubbed the real-space trick.
Using the above defined form-factor-bonds we can rewrite the projections explicitly starting with the $C$ to $P$ projection using $\bvec q_C = \bvec k_1+\bvec k_3-\bvec q_P$, see Eq.~\eqref{eq:simple_proj} till Eq.~\eqref{eq:adv_proj}.

\begin{figure*}[!hbt]
\begin{align}
\hat{P}[\hat{C}^{-1}[C]&]_{o_1,o_3}^{b_1,b_3}(\bvec q_P)_{s_1;s_3}^{s_2;s_4} = \int d\bvec k_1\, d\bvec {k}_3\, \sum_{o_2,o_4} g_{b_1}(o_2,\bvec k_1)g^*_{b_3}(o_4,\bvec k_3)\hat{C}^{-1}[C]_{o_1, o_2;o_3, o_4}(\bvec k_1,\bvec k_2;\bvec k_3)_{s_1;s_3}^{s_2;s_4} \label{eq:simple_proj}
\\
&= \int d\bvec k_1\, d\bvec {k}_3\, e^{-i\bvec k_1\bvec B_1}  \delta_{\bvec r_{1}+\bvec b_1,\bvec r_{2}}e^{i\bvec k_3\bvec B_3}  \delta_{\bvec r_{3}+\bvec b_3,\bvec r_{4}} \nonumber
 e^{i\bvec k_1\bvec B_1'} \delta_{\bvec r_{1}+\bvec b_1',\bvec r_{4}}e^{-i\bvec k_3\bvec B_3'}  \\
 &\qquad \qquad \delta_{\bvec r_{3}+\bvec b_3',\bvec r_{2}}C_{o_1,o_3}^{b_1',b_3'}(\bvec q_C)_{s_1;s_3}^{s_4;s_2}\\
&=
 \int d\bvec k_1\, d\bvec {k}_3\, \sum_{b_1',b_3'} e^{-i\bvec k_1(\bvec B_1-\bvec B_1')}  e^{i\bvec k_3 (\bvec B_3-\bvec B_3')} \nonumber
\delta_{\bvec r_{3}+\bvec b_3',\bvec r_{1}+\bvec b_1}\delta_{\bvec r_{1}+\bvec b_1',\bvec r_{3}+\bvec b_3}\\
&\qquad\qquad\qquad \sum_{\bvec R}e^{i\bvec q_C \bvec R}C_{o_1,o_3}^{b_1',b_3'}(\bvec R)_{s_1;s_3}^{s_4;s_2}
 \\
&= \int d\bvec k_1\, d\bvec k_3\, \sum_{b_1',b_3'} e^{-i\bvec k_1(\bvec B_1-\bvec B_1')}  e^{i\bvec k_3 (\bvec B_3-\bvec B_3')} 
\delta_{\bvec r_{3}+\bvec b_3',\bvec r_{1}+\bvec b_1}\delta_{\bvec r_{1}+\bvec b_1',\bvec r_{3}+\bvec b_3}\nonumber \\
&\qquad\qquad\qquad \sum_{\bvec R}e^{i( \bvec k_1+\bvec k_3-\bvec q_P) \bvec R}C_{o_1,o_3}^{b_1',b_3'}(\bvec R)_{s_1;s_3}^{s_4;s_2}
\\
&= \sum_{b_1',b_3',\bvec R} \delta_{\bvec R,\bvec B_1-\bvec B_1'}\delta_{\bvec R,\bvec B_3'-\bvec B_3} 
\delta_{\bvec r_{3}+\bvec b_3',\bvec r_{1}+\bvec b_1}\delta_{\bvec r_{1}+\bvec b_1',\bvec r_{3}+\bvec b_3}
 e^{-i\bvec q_P \bvec R} \int d \bvec q_C \, e^{-i\bvec q_C \bvec R}C_{o_1,o_3}^{b_1',b_3'}(\bvec q_C)_{s_1;s_3}^{s_4;s_2}
 \\
&= \int d \bvec q_C \;\sum_{b_1',b_3'}\delta_{\bvec B_3'-\bvec B_3,\bvec B_1-\bvec B_1'}
\delta_{\bvec r_{3}+\bvec b_3',\bvec r_{1}+\bvec b_1}\delta_{\bvec r_{1}+\bvec b_1',\bvec r_{3}+\bvec b_3}
 e^{i\bvec q_P (\bvec B_3-\bvec B_3')} e^{i\bvec q_C (\bvec B_3-\bvec B_3')}C_{o_1,o_3}^{b_1',b_3'}(\bvec q_C)_{s_1;s_3}^{s_4;s_2}.
 \label{eq:adv_proj}
\end{align}
\end{figure*}
The idea is to perform a Fourier transformation of $C$ in order to decouple the integration over momenta from the channel.
In a second step, we then revert this transformation  obtaining a more compact description of the formulas.
The projections can now be implemented as follows:
For each combination of incoming indices and bonds $(o_1,b_1,o_3,b_3)$ (TO) we store the allowed index combinations $(o_1,b_1',o_3,b_3')$ (FROM).
For each element TO we need to store the offset to the first corresponding element in the FROM list, as well as the number of elements corresponding to this.
Additionally, we cache all occurring $e^{i\bvec q_C (\bvec B_3-\bvec B_3')}$, where $(\bvec B_3-\bvec B_3')$ has to be mapped back to the minimal representation within the extended unit cell, we will call this the form-factor map.
Lastly, we need to store for each of the FROM and TO elements, which element of the form-factor map corresponds to it. In total we thus have five distinct arrays for the projection, which is reduced to a highly sparse reordering plus multiplication with a prefactor.
The other projections can be treated analogously and the derivations can be found in App.~\ref{App::proj}. An advantage of this rewriting is not only that we removed one momentum integration, but also do not sum over four independent form factors-bonds. The Kronecker-deltas remove one of the summations, thus effectively reducing the scaling to $\propto N_b^3$. In total, each of the inter-channel projections scales $\propto N_s^2N_{\sigma}^4N_k^2N_b^3$, which clearly is better than the initial $\propto N_s^2N_{\sigma}^4N_k^3N_b^4$

\section{Advantages and Limitations}
\begin{figure*}[!hbt]
    \centering
    \includegraphics[width = 0.32\linewidth]{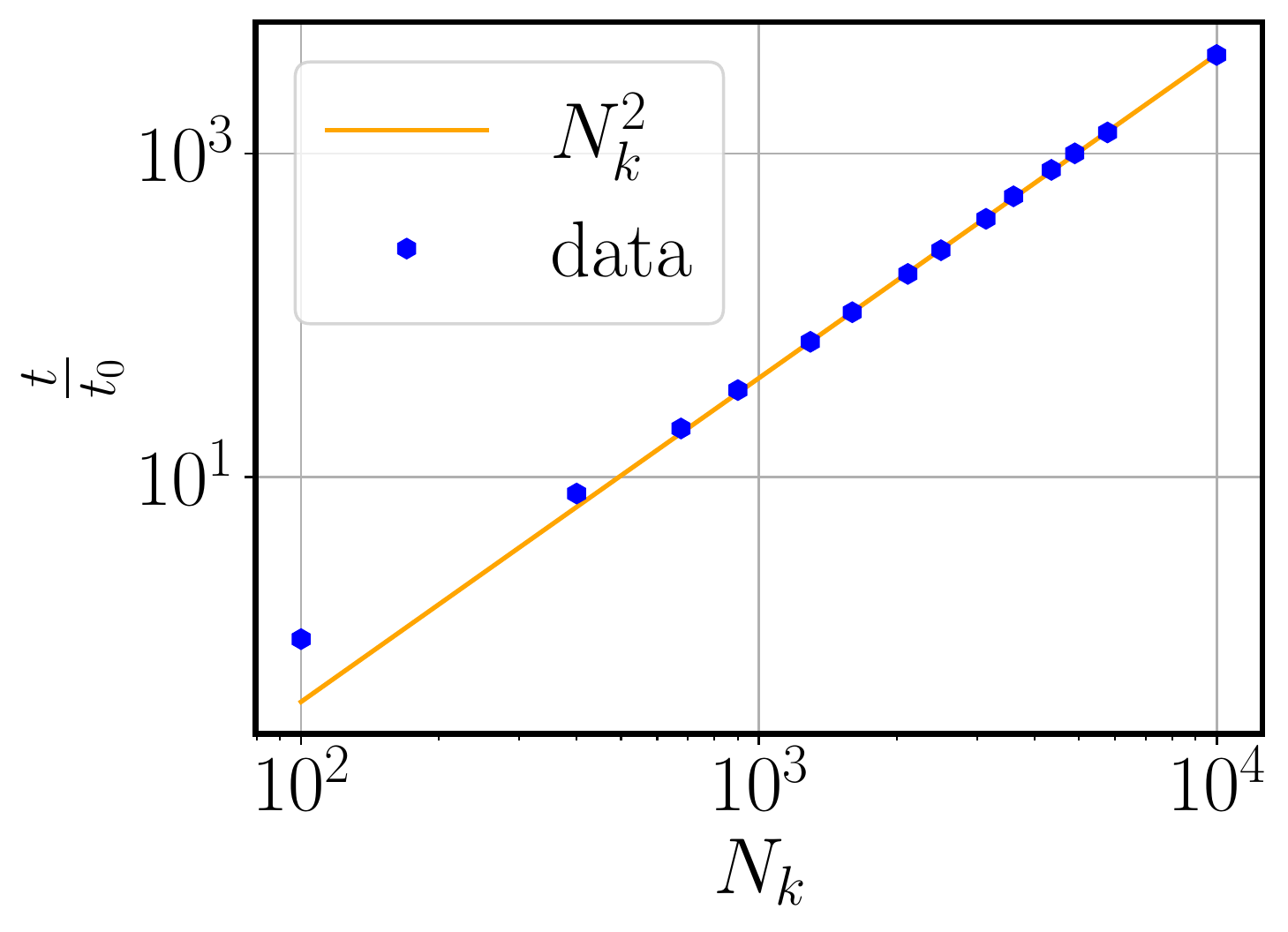}
    \includegraphics[width = 0.32\linewidth]{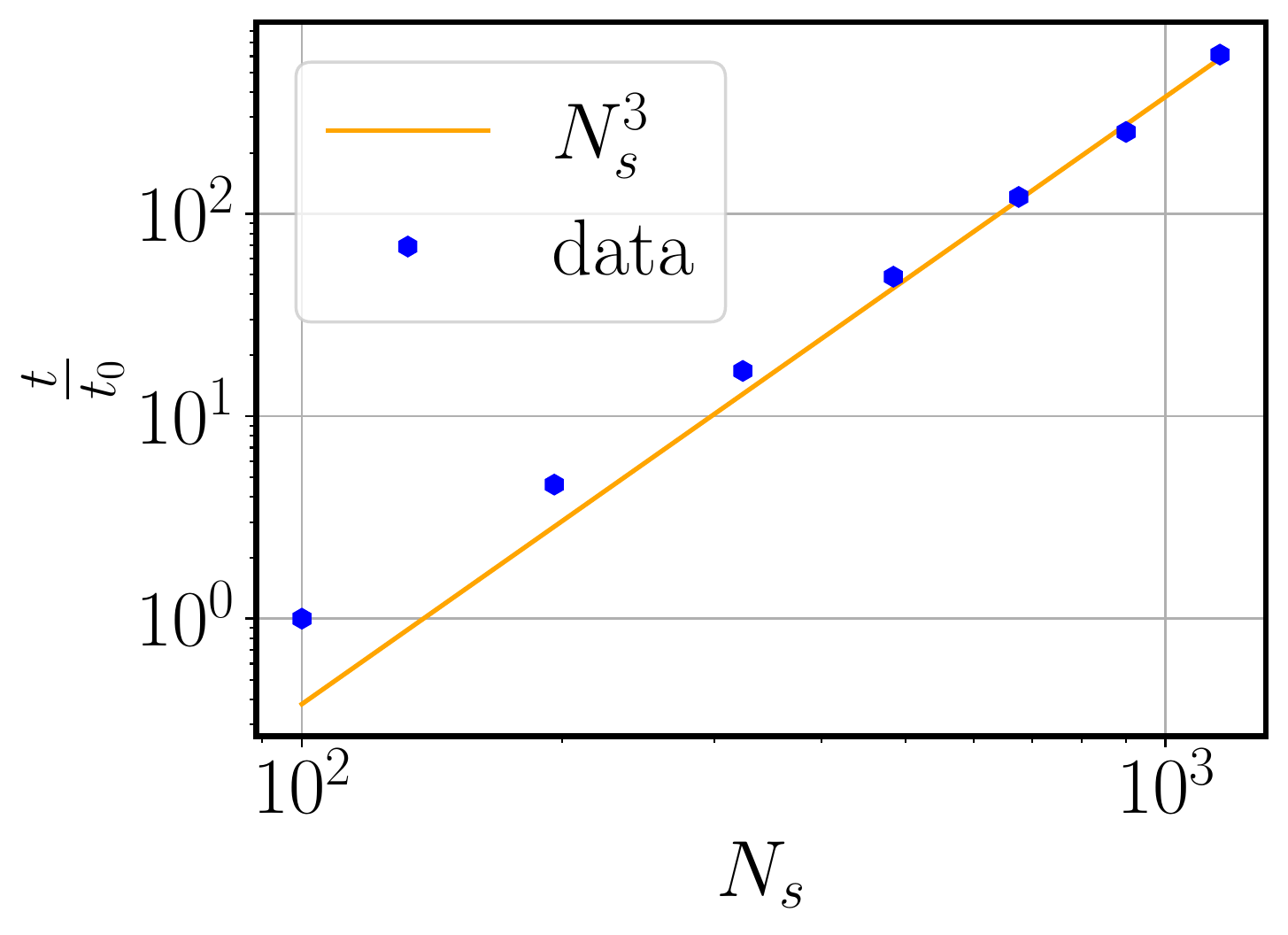}
    \includegraphics[width = 0.32\linewidth]{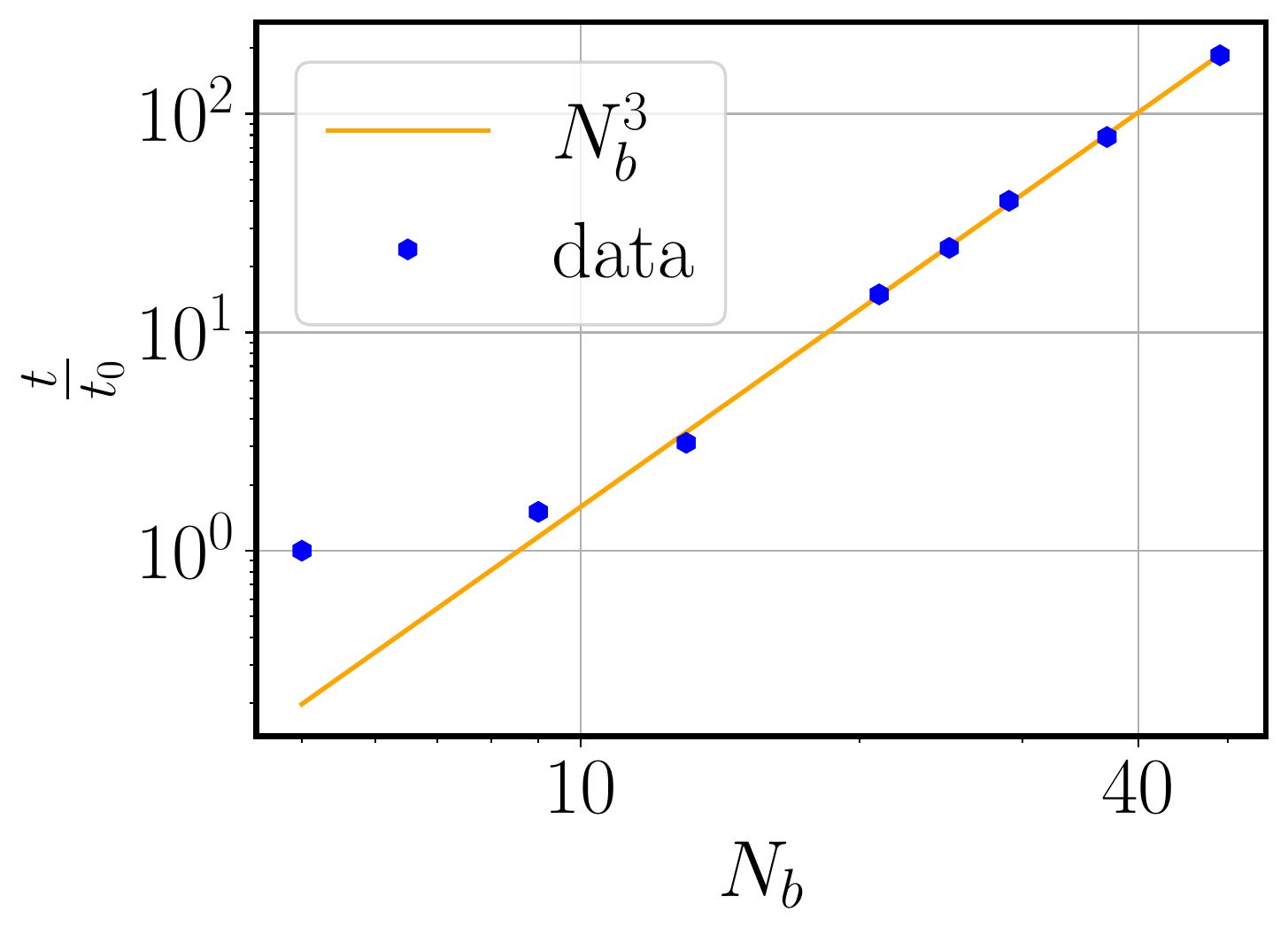}
    \caption{Scaling of the time $t$ needed to perform a single flow step with respect to the number of momentum points (left), number of sites (center) and number of bonds, right. The scaling expected from the analytical expressions is shown in orange.}
    \label{fig:scaling}
\end{figure*}
Here we will shortly discuss the main advantages and limitations of this approach.
The additional calculations required for the TU reduce the scaling, but increase the time constant of the calculation. Therefore, it can be numerically more efficient to use a grid FRG implementation for models with small unit cells, the square lattice Hubbard model to name just one.
Another drawback is the reliance on short ranged interactions, as the method is based on only developing long ranged behavior at high orders in $U$. This is however not as drastic as naively expected~\cite{lichtenstein_functional_2018}. {Furthermore, this restriction to short ranged interaction introduces a slight bias towards short ranged fluctuations as inter-channel contributions are restricted to certain length scales. In form-factor-bond converged calculations this bias is however negligible.}
The advantages are superior scaling in all degrees of freedom, allowing for computations in many models previously unaccessible to FRG. 
The main scalings of a flow step are shown in Fig.~\ref{fig:scaling}. Using hybrid architecture and MPI, the time constant can be brought down further. Especially the usage of GPUs can reduce the the computation time a lot. To verify the validity of the implementation we ensured the reproduction of FRG benchmark results~\cite{in-prep-beyer}.

In stark contrast to brute force FRG, this implementation is not memory bound anymore, instead we are bound for most use cases by the computation time due to the linear scaling of the memory in the number of momentum points.
The exception to this rule are systems with very large unit cells. Another big advantage of the method presented here, is that it can be readily employed in models without translational symmetry. We simply have to leave the momentum variables out of the equations and arrive at a real-space TUFRG formalism~\cite{hauck_electronic_2021}.

\begin{figure}[!tbh]
    \centering
    \includegraphics[width = 0.99\linewidth]{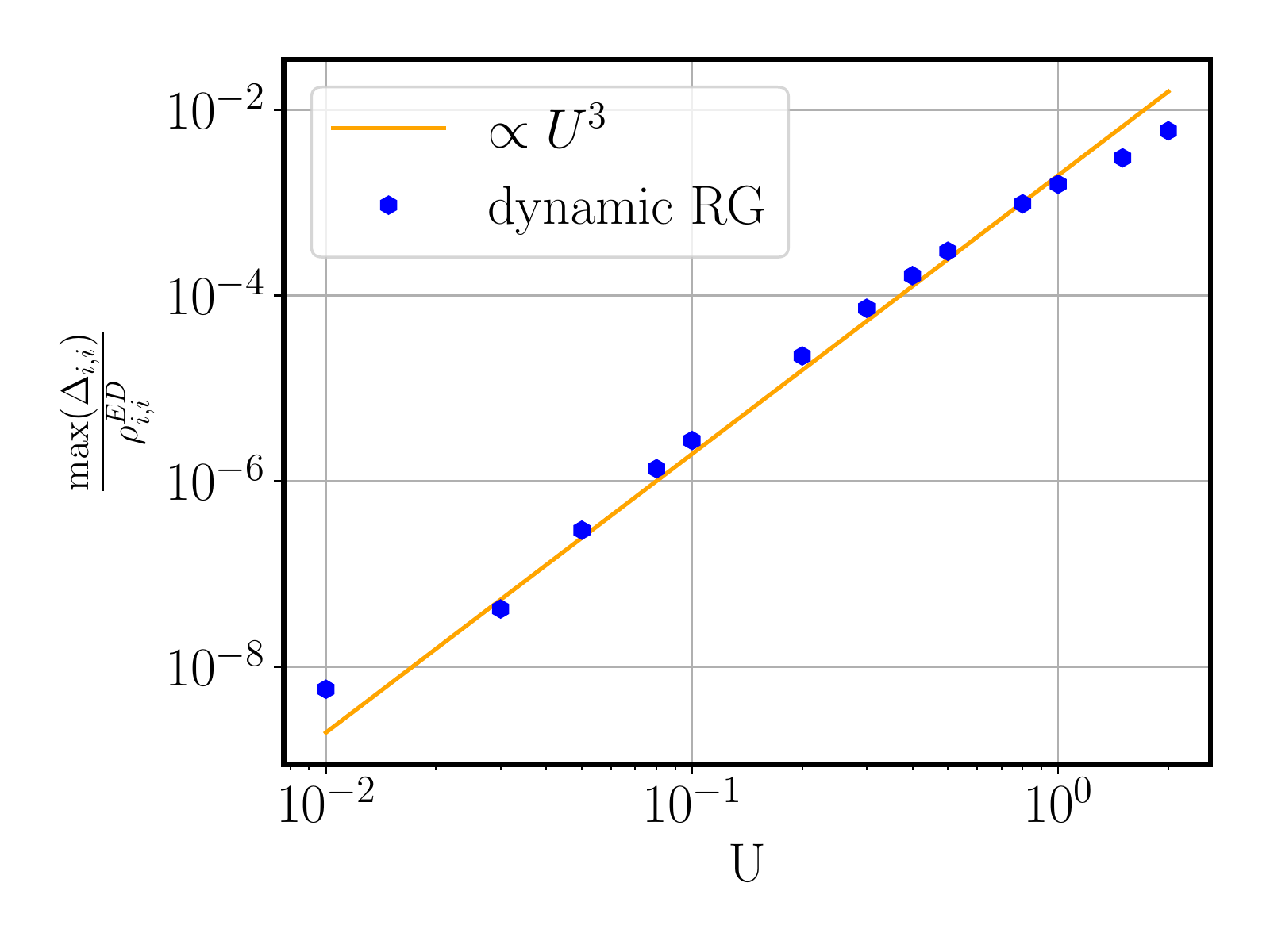}
    \caption{Relative error of the occupation number predicted by the single frequency TU$^2$FRG compared to exact diagonalization. Calculations were perfomed at $T=0$ an $8$ site open boundary conditions Hubbard chain with random on-site potentials. Even at intermediate interactions, here $U=2$, the relative the error does not exceed $1\%$}
    \label{fig:error_vs_ed}
\end{figure}

{
\section{Application}
Many body localization~\cite{2006-AA,2010,Nandkishore_MBL_2015,2016}  has been a central topic in condensed matter theory in recent years.
Its observation in optical gases~\cite{schreiber2015observation,choi2016exploring} lead to a surge in theoretical works trying to understand this phenomenon.
However, the methods applicable in this case are limited as the thermalization hypothesis does not hold  and the systems are non-translationally invariant.
So far there mostly DMRG~\cite{decker2021manybody,lim2016many} and ED~\cite{2010} have been applied to investigate these phenomena. Both methods do not scale favorable in 2D. Here we aim to show, that the presented method could be used to study this phenomenon. For this purpose we consider a $8$ site Hubbard chain with open boundary conditions, a random on site potential and on-site interactions
\begin{align}
 \centering
     H = -\sum_{ i,j, \sigma} (t\delta_{<i,j>} &+ r_i \delta_{i,j})c^{\dagger}_{ i,\sigma} c_{ j,\sigma} + \sum_{ i} U n_{ i, \uparrow} n_{ i, {\downarrow}},\label{eq:Ham}
\end{align}
where $\delta_{<i,j>}=1$ if $i$ and $j$ are neighbouring sites, $t = 1$ is chosen as unit of energy and $r_i$ is a random number  $\in [-0.5,0.5]$.
As we include all terms up to order $U^3$, we expect the error to scale accordingly, which is verified in Fig.~\ref{fig:error_vs_ed}. 

Even though we do not incorporate the full frequency content we stay below {a} relative error of about $1\%$ up to $U=2$. This offers a route to further exploring this fascinating phenomenon in higher dimensions with FRG.}

\section{Conclusion and Outlook}
We presented a new full unity-space derivation of the level-two truncated FRG flow equations up to first loop order. The real-space variant of this approach has already been successfully applied to quasicrystals~\cite{hauck_electronic_2021} and finite sized models~\cite{hauck_strong_2021}.
The TU$^2$FRG we derived further extends the grasp of FRG significantly and enables calculations for many interesting systems. Additionally, we presented possible implementation strategies for the key operations of the flow of $\Gamma^4$ and have shown that the optimal scalings can be reached. {Furthermore, we showed that the present implementation could be used to investigate disorder effects and possible many-body localization on a qualitative level.}

The next step towards establishing FRG as the go to method for electronic instability calculations at weak to intermediate couplings is to prove its applicability in systems of interest, such as multi-layer graphene~\cite{Zhou-2021-half,Zhou-2021-supercon,zhou2021isospin}, twisted materials~\cite{Can2021-TCup,cao_unconventional_2018} and Kagome metals~\cite{Ye_2018_kagome_metal}.
Furthermore, it is desirable to implement a more sophisticated frequency unity which then enables us to study phonon and photon mediated superconductivity. Recent works in this direction~\cite{yirga_frequency_2021} show promising results. {The combination with the recently developed single boson exchange formulation of the FRG~\cite{bonetti2021single} also is an interesting route for further developments. The derived formalism is also directly applicable in Parquet approaches~\cite{eckhardt_truncated_2020}, extending the applicability of these method to larger unit cell models.}

\begin{acknowledgements}
We thank J.~Beyer, C.~Honerkamp and L.~Klebl for fruitful discussions and comparison of results. The Deutsche Forschungsgemeinschaft (DFG, German Research Foundation) is acknowledged for support through RTG 1995 and under Germany's Excellence Strategy - Cluster of Excellence Matter and Light for Quantum Computing (ML4Q) EXC 2004/1 - 390534769. We acknowledge support from the Max Planck-New York City Center for Non-Equilibrium Quantum Phenomena. Simulations were performed with computing resources granted by RWTH Aachen University under project rwth0742.
\end{acknowledgements}
\section{Authors contributions}
JBH performed the derivations, implemented the code and ran the simulations. DMK supervised the work.
All the authors were involved in the preparation of the manuscript. \\ \\
{\textbf{Data availability} The datasets generated during and/or analysed during the current study are available from the corresponding author on reasonable request.}
\section*{Appendix}
\appendix
\section{Flow equations}
\label{App::floweq}
In the following we derive the non-$SU(2)$ invariant flow equations for the $C$ and $D$ projected channels. We begin with the $C$ projected channel starting from Eq.~\eqref{eq:Cflow_beg} and with the $D$-channel starting from Eq.~\eqref{eq:Dflow_beg}. 
\begin{figure*}
\begin{align}
\frac{{\rm d}\hat{C}[\Phi^C]_{o_1,o_3}^{b_1,b_3}(q_C)_{s_1;s_3}^{s_4;s_2}}{{\rm d}\Lambda} &=   g_{b_1}(o_4, k_1)g^*_{b_3}(o_2, k_3) \Gamma_{o_1, o_4';o_1', o_4}^{s_1,s_4';s_1',s_4}( k_1,k_1-q_C; p)\nonumber
\\ 
&\qquad\quad  \cdot \left(G_{o_1';o_3'}^{s_1',s_3'}({ p})S_{o_4';o_2'}^{s_4',s_2'}(p-q_C) + G\leftrightarrow S) \right) \Gamma_{o_3', o_2;o_3, o_2'}^{s_3',s_2;s_3,s_2'}(p, p-q_C; k_3)\label{eq:Cflow_beg}
\\
&\quad= \nonumber g_{b_1}(o_4, k_1)g^*_{b_3}(o_2, k_3)\Gamma_{o_1, o_4';o_1', o_4}^{s_1,s_4';s_1',s_4}( k_1,k_1-q_C; p)
\\
&\qquad\quad \nonumber\cdot \delta( p- p_1)\delta_{o_4',n_4}\left(G_{o_1';o_3'}^{s_1',s_3'}({ p_1})S_{n_4;n_2}^{s_4',s_2'}(p_1-q_C) + G\leftrightarrow S) \right) 
\\
&\qquad\quad \cdot \delta( p_2- p_1) \delta_{o_2',n_2} \Gamma_{o_3', o_2;o_3, o_2'}^{s_3',s_2;s_3,s_2'}(p, p_2-q_C; k_3)
\\
&\quad=   \nonumber g_{b_1}(o_4, k_1) g^*_{b_1'}(o_4', p) \Gamma_{o_1, o_4';o_1', o_4}^{s_1,s_4';s_1',s_4}( k_1,k_1-q_C; p)
\\ 
& \qquad\quad \nonumber \cdot g_{b_1'}(n_4, p_1)g^*_{b_3'}(n_2, p_1)\left(G_{o_1';o_3'}^{s_1',s_3'}({ p_1})S_{n_4;n_2}^{s_4',s_2'}(p_1-q_C) + G\leftrightarrow S) \right) 
\\ 
& \qquad\quad \cdot  g_{b_3'}(o_2', p_2) \Gamma_{o_3', o_2;o_3, o_2'}^{s_3',s_2;s_3,s_2'}(p, p_2-q_C; k_3)g^*_{b_3}(o_2, k_3) 
\\
&\quad= \hat{C}[\Gamma]_{o_1,o_1'}^{b_{1},b_{1}'}( q_C)_{s_1;s_1'}^{s_4;s_4'}\,L^{ph;b_{1}',b_{3}'}_{o_1',o_3'}( q_C)_{s_1';s_3'}^{s_4';s_2'}\hat{C}[\Gamma]_{o_3',o_3}^{b_{3}',b_{3}}( q_C)_{s_3';s_3}^{s_2';s_2}.
\label{eq:Cflow_end}
\end{align}
\end{figure*}

\begin{figure*}
\begin{align}
\frac{{\rm d}\hat{D}[\Phi^D]_{o_1,o_4}^{b_1,b_4}(q_D)_{s_1;s_4}^{s_3;s_2}}{{\rm d}\Lambda} &=  - g_{b_1}(o_3, k_1)g^*_{b_4}(o_2, k_4) \Gamma_{o_1, o_4';o_3, o_1'}^{s_1,s_4';s_3,s_1'}( k_1,p-q_D; k_1-q_D)\nonumber 
\\ 
&\qquad\quad   \cdot \left(G_{o_1';o_3'}^{s_1',s_3'}({ \label{eq:Dflow_beg}  p-q_D})S_{o_4';o_2'}^{s_4',s_2'}(p) + G\leftrightarrow S) \right) \Gamma_{o_3', o_2;o_2', o_4}^{s_3',s_2;s_2',s_4}(p, k_4-q_D; p-q_D)
\\
&\quad=\nonumber - g_{b_1}(o_3, k_1)g^*_{b_4}(o_2, k_4) \Gamma_{o_1, o_4';o_3, o_1'}^{s_1,s_4';s_3,s_1'}( k_1,p-q_D; k_1-q_D)
\\
&\qquad\quad \nonumber \cdot \delta( p- p_1)\delta_{o_4',n_4}\left(G_{o_1';o_3'}^{s_1',s_3'}({ p_1-q_D})S_{n_4;n_2}^{s_4',s_2'}(p_1) + G\leftrightarrow S) \right) 
\\
&\qquad\quad \cdot \delta( p_2- p_1) \delta_{o_2',n_2} \Gamma_{o_3', o_2;o_2', o_4}^{s_3',s_2;s_2',s_4}(p, k_4-q_D; p-q_D)
\\
&\quad= \nonumber -g_{b_1}(o_3, k_1) g^*_{b_1'}(o_4', p)  \Gamma_{o_1, o_4';o_3, o_1'}^{s_1,s_4';s_3,s_1'}( k_1,p-q_D; k_1-q_D)
\\ 
& \qquad\quad \nonumber  \cdot g^*_{b_3'}(n_2, p_1)g_{b_1'}(n_4, p_1)\left(G_{o_1';o_3'}^{s_1',s_3'}({ p_1-q_D})S_{n_4;n_2}^{s_4',s_2'}(p_1) + G\leftrightarrow S) \right) 
\\ 
& \qquad\quad   \cdot g_{b_3'}(o_2', p_2)g^*_{b_4}(o_2, k_4) \Gamma_{o_3', o_2;o_2', o_4}^{s_3',s_2;s_2',s_4}(p_2, k_4-q_D; p_2-q_D)
\\
&\quad= -\hat{D}[\Gamma]_{o_1,o_1'}^{b_{1},b_{1}'}( q_D)_{s_1;s_1'}^{s_3;s_4'}L^{ph;b_1',b_{3}'}_{o_1',o_3'}( q_D)_{s_1';s_3'}^{s_4';s_2'}\hat{D}[\Gamma]_{o_3',o_4}^{b_{3}',b_{4}}( q_D)_{s_3';s_4}^{s_2';s_2}.\label{eq:Dflow_end}
\end{align}
\end{figure*}

In the more specialized, but very regularly used case of $SU(2)$-invariant models, we can simplify these flow equations by explicitly enforcing the crossing relations (Eq.~\eqref{eq:crossrel}). Additionally, in this case the Hamiltonian is spin conserving, i.e.~$H_{s_1,s_2} \propto \delta_{s_1,s_2}$, simplifying the loop derivatives to being spin diagonal. 
\begin{align}
    \Gamma_{o_1.._4}^{s_1.._4}(k_1.._3) = \nonumber &V_{o_1,o_2,o_3,o_4}(k_1,k_2,k_3)\delta_{s_1,s_3}\delta_{s_2,s_4}\\ -  &\tilde{V}_{o_1,o_2,o_4,o_3}(k_1,k_2,k_4)\delta_{s_1,s_4}\delta_{s_2,s_3} 
    \label{eq:crossrel}
\end{align}
The crossing relations follow from the transformation of the effective action into the irreducible representations in spin space~\cite{honerkamp-salmhofer-2001}. For the full vertex we find that $V=\tilde{V}$. Thus, instead of performing the flow for the spin dependent vertex function, we can perform it for the spin independent vertex $V$, sparing us a computational complexity of $N_s^6$. Again we can split the flow into three different channels, but special care has to be taken of the $C$ and $D$ channel, as here we find $\tilde{V}_C = V_D$ and vice versa. The flow equations for the channel specific $V^X$-functions can now be obtained by picking a specific spin combination, for example $\uparrow,\downarrow;\downarrow,\uparrow$. The flow equations are then adopted analytically by inserting the decomposition from Eq.~\eqref{eq:crossrel} for $\Gamma$, as can be seen in Eq.~\eqref{eq:proj_P_flowSU}.

\begin{align}
\frac{{\rm d}\hat{P}[V^P]_{o_1,o_3}^{b_{1},b_3}( q_P)}{{\rm d}\Lambda}&= \frac{1}{2} \hat{P}[\Gamma]_{o_1,o_1'}^{b_{1},b_{1}'}( q_P)_{\uparrow,\downarrow;s_1',s_2'}\nonumber
\\
 \cdot L^{pp;b_{1}',b_{3}'}_{o_1',o_3'}&( q_P)\hat{P}[\Gamma]_{o_3',o_3}^{b_{3}',b_{3}}( q_P)_{s_1',s_2';\downarrow,\uparrow}\\
=-\hat{P}[V]_{o_1,o_1'}^{b_{1},b_{1}'}&( q_P) L^{pp;b_{1}',b_{3}'}_{o_1',o_3'}( q_P)\hat{P}[V]_{o_3',o_3}^{b_{3}',b_{3}}( q_P)
\label{eq:proj_P_flowSU}
\end{align}

\begin{align}
&\frac{{\rm d}\hat{C}[V^C]_{o_1,o_3}^{b_{1},b_3}( q_C)}{{\rm d}\Lambda} = \nonumber
 -\hat{C}[V]_{o_1,o_1'}^{b_{1},b_{1}'}( q_C) \\ &\qquad\qquad\qquad\quad\cdot L^{ph;b_{1}',b_{3}'}_{o_1',o_3'}( q_C) \hat{C}[V]_{o_3',o_3}^{b_{3}',b_{3}}( q_C), 
\\
&\frac{{\rm d}\hat{D}[V^D]_{o_1,o_3}^{b_{1},b_3}( q_D)}{{\rm d}\Lambda} = \nonumber 2\left[\hat{D}[V]-\frac{\hat{C}[V]}{2}\right]_{o_1,o_1'}^{b_{1},b_{1}'}( q_D) \\ &\qquad\qquad\qquad\quad\cdot L^{ph;b_{1}',b_{3}'}_{o_1',o_3'}( q_D)\nonumber\left[\hat{D}[V]-\frac{\hat{C}[V]}{2}\right]_{o_3',o_3}^{b_{3}',b_{3}}(q_D)
\\&\qquad\qquad\qquad\quad +\frac{{\rm d}\hat{C}[V^C]_{o_1,o_3}^{b_{1},b_3}( q_D)}{2d\Lambda}.
\end{align}
Here we already reduced the number of vertex-loop-vertex contractions for the $D$-channel from three to one by performing a completion of the square. This has proven to be crucial in the case of large unit cells as we basically reduce the computational effort by $\frac{2}{5}$.

\section{Projections}
\label{App::proj}
To enable the reader to directly start implementing its own TU$^2$FRG framework, we give in the following the rest of the inter-channel projections. The derivations are quite lengthy and therefore we again use our modified summing convention. We only need to derive four inter-channel projections, as the $C$ to $P$ and the $P$ to $C$ projection are the same, as well as the projections between the $C$ and $D$ channel. The implementation strategy is the same as described in Subsec.~\ref{ssec:proj}

\begin{figure*}[!bht]
\begin{align}
\hat{P}[\hat{D}^{-1}[D]]_{o_1,o_3}^{b_1,b_3}(\bvec q_P)_{s_1;s_3}^{s_2;s_4} &=  e^{-i\bvec k_1\bvec B_1}  \delta_{\bvec r_{1}+\bvec b_1,\bvec r_{2}}e^{i\bvec k_3\bvec B_3}  \delta_{\bvec r_{3}+\bvec b_3,\bvec r_{4}} 
 e^{i\bvec k_1\bvec B_1'} \delta_{\bvec r_{1}+\bvec b_1',\bvec r_{3}}e^{-i(\bvec q_P-\bvec k_3)\bvec B_4'} 
\delta_{\bvec r_{4}+\bvec b_4',\bvec r_{2}}D_{o_1,o_4}^{b_1',b_4'}(\bvec q_D)_{s_1;s_4}^{s_3;s_2}\\
&= e^{-i\bvec k_1(\bvec B_1-\bvec B_1')}  e^{i\bvec k_3 (\bvec B_4'+\bvec B_3)} e^{-i\bvec q_P\bvec B_4'} 
\delta_{\bvec r_{1}+\bvec b_1,\bvec r_{3}+\bvec b_3+\bvec b_4'}\delta_{\bvec r_{1}+\bvec b_1',\bvec r_{3}}
e^{i(\bvec k_1-\bvec k_3) \bvec R}D_{o_1,o_3+b_3}^{b_1',b_4'}(\bvec R)_{s_1;s_4}^{s_3;s_2}
 \\
&= \delta_{\bvec B_1-\bvec B_1', \bvec R}  \delta_{\bvec B_4'+\bvec B_3,\bvec R} e^{-i\bvec q_P\bvec B_4'} 
\delta_{\bvec r_{1}+\bvec b_1,\bvec r_{3}+\bvec b_3+\bvec b_4'}\delta_{\bvec r_{1}+\bvec b_1',\bvec r_{3}}
e^{-i \bvec q_D \bvec R}D_{o_1,o_3+b_3}^{b_1',b_4'}(\bvec q_D)_{s_1;s_4}^{s_3;s_2}
 \\
&= \delta_{\bvec B_1-\bvec B_1', \bvec B_4'+\bvec B_3}  
\delta_{\bvec r_{1}+\bvec b_1,\bvec r_{3}+\bvec b_3+\bvec b_4'}\delta_{\bvec r_{1}+\bvec b_1',\bvec r_{3}}e^{-i\bvec q_P\bvec B_4'} e^{-i \bvec q_D (\bvec B_1-\bvec B_1')}
D_{o_1,o_3+b_3}^{b_1',b_4'}(\bvec q_D)_{s_1;s_4}^{s_3;s_2}
 \label{eq:PD_adv_proj}
\end{align}

\begin{align}
\hat{C}[\hat{D}^{-1}[D]]_{o_1,o_3}^{b_1,b_3}(\bvec q_C)_{s_1;s_3}^{s_4;s_2} &=  e^{-i\bvec k_1\bvec B_1}  \delta_{\bvec r_{1}+\bvec b_1,\bvec r_{4}}e^{i\bvec k_3\bvec B_3}  \delta_{\bvec r_{3}+\bvec b_3,\bvec r_{2}} 
 e^{i\bvec k_1\bvec B_1'} \delta_{\bvec r_{1}+\bvec b_1',\bvec r_{3}}e^{-i(\bvec k_1-\bvec q_C)\bvec B_4'} 
\delta_{\bvec r_{4}+\bvec b_4',\bvec r_{2}}D_{o_1,o_4}^{b_1',b_4'}(\bvec q_D)_{s_1;s_4}^{s_3;s_2}\\
&=  e^{-i\bvec k_1(\bvec B_1+\bvec B_4'-\bvec B_1')} e^{i\bvec k_3\bvec B_3}
\delta_{\bvec r_{1}+\bvec b_1',\bvec r_{3}}
\delta_{\bvec r_{1}+\bvec b_1+\bvec b_4',\bvec r_{3}+\bvec b_3}e^{i\bvec q_C\bvec B_4'}e^{i(\bvec k_1-\bvec k_3) \bvec R} D_{o_1,o_1+b_1}^{b_1',b_4'}(\bvec R)_{s_1;s_4}^{s_3;s_2}\\
&=  \delta_{\bvec B_1+\bvec B_4'-\bvec B_1',\bvec R} \delta_{\bvec B_3,\bvec R}
\delta_{\bvec r_{1}+\bvec b_1',\bvec r_{3}}
\delta_{\bvec r_{1}+\bvec b_1+\bvec b_4',\bvec r_{3}+\bvec b_3}e^{i\bvec q_C\bvec B_4'}e^{i\bvec q_D \bvec R} D_{o_1,o_1+b_1}^{b_1',b_4'}(\bvec q_D)_{s_1;s_4}^{s_3;s_2}\\
&=  \delta_{\bvec B_1+\bvec B_4'-\bvec B_1',\bvec B_3}
\delta_{\bvec r_{1}+\bvec b_1',\bvec r_{3}}
\delta_{\bvec r_{1}+\bvec b_1+\bvec b_4',\bvec r_{3}+\bvec b_3}e^{i\bvec q_C\bvec B_4'}e^{-i\bvec q_D \bvec B_3} D_{o_1,o_1+b_1}^{b_1',b_4'}(\bvec q_D)_{s_1;s_4}^{s_3;s_2}
 \label{eq:CD_adv_proj}
\end{align}

\begin{align}
\hat{D}[\hat{P}^{-1}[P]]_{o_1,o_4}^{b_1,b_4}(\bvec q_D)_{s_1;s_4}^{s_3;s_2} &=  e^{-i\bvec k_1\bvec B_1}  \delta_{\bvec r_{1}+\bvec b_1,\bvec r_{3}}e^{i\bvec k_4\bvec B_4}  \delta_{\bvec r_{4}+\bvec b_4,\bvec r_{2}} 
 e^{i\bvec k_1\bvec B_1'} \delta_{\bvec r_{1}+\bvec b_1',\bvec r_{2}}e^{-i(\bvec k_1 - \bvec q_D)\bvec B_3'} 
\delta_{\bvec r_{3}+\bvec b_3,\bvec r_{4}}P_{o_1,o_3}^{b_1',b_3'}(\bvec q_P)_{s_1;s_3}^{s_2;s_4}\\
&=  e^{-i\bvec k_1(\bvec B_1-\bvec B_1'+\bvec B_3')}  e^{i\bvec k_4\bvec B_4}   \delta_{\bvec r_{1}+\bvec b_1',\bvec r_{4}+\bvec b_4} \delta_{\bvec r_{1}+\bvec b_1+\bvec b_3,\bvec r_{4}}
e^{i\bvec q_D\bvec B_3'} e^{i(\bvec k_1 + \bvec k_4- \bvec q_D)\bvec R} P_{o_1,o_1+b_1}^{b_1',b_3'}(\bvec R)_{s_1;s_3}^{s_2;s_4}\\
&=  \delta_{\bvec B_1-\bvec B_1'+\bvec B_3', \bvec R}  \delta_{-\bvec B_4, \bvec R}   \delta_{\bvec r_{1}+\bvec b_1',\bvec r_{4}+\bvec b_4} \delta_{\bvec r_{1}+\bvec b_1+\bvec b_3,\bvec r_{4}}
e^{i\bvec q_D(\bvec B_3'-\bvec R)}e^{-i\bvec q_P\bvec R} P_{o_1,o_1+b_1}^{b_1',b_3'}(\bvec q_P)_{s_1;s_3}^{s_2;s_4}\\
&=  \delta_{\bvec B_1'-\bvec B_1-\bvec B_3', \bvec B_4}  \delta_{\bvec r_{1}+\bvec b_1',\bvec r_{4}+\bvec b_4} \delta_{\bvec r_{1}+\bvec b_1+\bvec b_3,\bvec r_{4}}
e^{i\bvec q_D(\bvec B_3'+\bvec B_4)}e^{i\bvec q_P\bvec B_4} P_{o_1,o_1+b_1}^{b_1',b_3'}(\bvec q_P)_{s_1;s_3}^{s_2;s_4}
 \label{eq:DP_adv_proj}
\end{align}
\end{figure*}

\FloatBarrier

\printbibliography

\end{document}